\documentclass[useAMS,usenatbib]{mn2e}
\usepackage{epsfig}
\usepackage{amsmath}
\usepackage{longtable}
\usepackage{rotating}
\usepackage{epstopdf}

\newcommand {\aplt} {\ {\raise-.5ex\hbox{$\buildrel<\over\sim$}}\ } 

\title{Short-term variability of 10 trans-Neptunian objects}
\author[A Thirouin et al]
{A. Thirouin $^{1}$\thanks{E-mail:
thirouin@iaa.es},
J.L. Ortiz$^{1}$, A. Campo Bagatin$^{2,3}$, P. Pravec$^{4}$, N. Morales$^{1}$, 
\newauthor
O. Hainaut$^{5}$, R. Duffard$^{1}$ \\
$^{1}$Instituto de Astrof\'{\i}sica  de Andaluc\'{\i}a - CSIC, Apt
3004, 18008  Granada,  Spain\\
$^{2}$Departamento de Fisica, Ingenieria de Sistemas y teoria de la Se\~{n}al, Universidad de Alicante, PO Box 99, 03080 Alicante, Spain\\
$^{3}$Instituto de F\'{\i}sica Aplicada a las Ciencias y la Tecnolog\'{\i}a, Universidad de Alicante, PO Box 99, 03080 Alicante, Spain\\
$^{4}$Astronomical Institute, Academy of Sciences of the Czech Republic, Fri\v{c}ova 1, CZ-25165 Ond\v{r}ejov, Czech Republic.\\
$^{5}$ESO, Karl-Schwarzschild-Str. 2, 85748 Garching bei M\"{u}nchen, Germany.\\
}
\begin{document}

\date{}

\pagerange{\pageref{firstpage}--\pageref{lastpage}} \pubyear{2012}

\maketitle

\label{firstpage}

\begin{abstract}
We present our latest results about the short-term variability of trans-Neptunian objects
(TNOs). We performed broad-band CCD photometric observations using several telescopes
in Spain and Chile. We present results based on three years of observations and report the
short-term variability of 10 TNOs. Our sample of studied targets contains classical objects: (275809) 2001~QY$_{297}$, (307251) 2002~KW$_{14}$, (55636) 2002~TX$_{300}$, 2004~NT$_{33}$, (230965) 2004~XA$_{192}$, and (202421) 2005~UQ$_{513}$, a resonant body: (84522) 2002~TC$_{302}$, a scattered target: (44594) 1999~OX$_{3}$, and two detached objects: (145480) 2005~TB$_{190}$, and (40314) 1999~KR$_{16}$.  
For each target, light curves as well as possible rotation periods and photometric amplitudes
are reported. The majority of the observed objects present a low peak-to-peak amplitude, $<$0.15~mag. Only two objects exhibit light curve amplitudes higher than 0.15~mag: (275809) 2001~QY$_{297}$, and (307251) 2002~KW$_{14}$. We remark two biases in the literature, previously studied in Thirouin et al. and confirmed by this new study: a bias towards objects with a small amplitude light curve and a second one against objects with a long rotational period in the data base of published rotational periods. We derived constraints on physical properties of some targets. We also report the solar phase curves of (40314) 1999~KR$_{16}$, and (44594) 1999~OX$_{3}$ for solar phase angles from 0$^{\circ}$ to around 2$^{\circ}$. Part of our discussion is focused on the study of (275809) 2001~QY$_{297}$ which turned out to be an asynchronous binary system. 

\end{abstract}


\section{Introduction}

The Edgeworth-Kuiper belt objects, usually called trans-Neptunian
objects (TNOs) orKuiper belt objects (KBOs), are known to be wellpreserved
fossil remnants of our Solar system formation. Since
the discovery of the first TNO (after Pluto) in 1992 by \cite{Jewitt1993}, various observational approaches to study the physical
properties of TNOs have been performed, including spectroscopic,
photometric and binarity studies. Our own approach to study these
objects is to detect the periodic variation of their brightness as a
function of time, resulting from their rotation \citep{Thirouin2010, Ortiz2007, Ortiz2006, Ortiz2004, Ortiz2003}. We analyse their rotational
periods, surfaces, shapes and internal structures studying their light
curves.    

Less than 5 per cent of the known TNOs have well-determined
rotational periods. Moreover, \cite{Sheppard2008} and \cite{Thirouin2010} pointed out that the sample of studied objects
is highly biased towards bright objects, large variability amplitudes and short rotational periods. Only 10 per cent of the rotational periods
published are larger than 10 h. The majority of light curve
amplitudes and rotational periods are published with large uncertainties
or, sometimes, they are just estimations or limiting values.
The sample of studied TNOs is essentially composed of bright
(visual magnitude $<$22 mag) and large objects. We can enumerate
various reasons in order to explain some of these biases. First,
we must point out observational limitations. A reliable study of
TNO rotational properties requires a lot of observational time on a
medium to large telescope. This causes a bias towards brighter objects,
but also short period and large amplitude.Another class of limitations
is due to reduction problems. A reliable photometric study
needs effective data reduction. Determining low-amplitude light
curves and/or detecting long rotation periods are very time consuming
and require a lot of observation time. Furthermore, 24-h aliases
frequently complicate the analysis of time-series photometry.

To help debias the sample of studied objects, longer term monitoring
is needed. This kind of observations is based on the coordination
of observational runs with various telescopes all around the world.
Using telescopes with similar characteristics in different continents
allows us to observe a target ‘continuously’. In other words, if we can monitor our targets during a long time, we can detect long rotation
periods and minimize the 24-h-aliases effect. In 2009 July,
we carried out our first coordinated campaign between Spain and
Chile. Part of this work presents results based on this coordinated
campaign for TNOs. Another part of this work is dedicated to our
programme on light curves of TNOs, started in 2001. In this work, we report our newest results based on observations carried out between
2008 and 2010. 

This paper is divided into six sections. Section 2 will describe
the observations and the data set analysed here. Section 3 will
describe our reduction techniques used in order to derive periods
and photometric ranges. In Section 4, we will give, for each target,
a summary of our main results. In Section 5, we will discuss our
results altogether. Section 6 is dedicated to the conclusion of this
work.

\section{Observations}
\subsection{Runs and Telescopes}
We present two different approaches to study the luminosity variability of TNOs. We analysed data obtained from the coordinated campaign and data from our regular programme on light curves of KBOs.

\subsubsection{Coordinated campaign}

In 2009 July, we carried out our first coordinated campaign involving
Europe and South America. Typically, an observational night of
July in Europe starts around 22 h UT and finishes at 5 h UT, whereas
an observational night in South America starts around 0 h UT and finishes at 10 h UT. Under perfect conditions and if the target is visible
in both sites during the entire night, we have five extra hours of
observational time. By using this approach, we have a continuous
time coverage of about 15 h, thereby addressing some of the biases
against long periods and the issue of the 24-h aliases. We carefully
coordinated the observations, to match exactly the field of view of
both telescopes, during the campaign. 

In Europe, we used the Telescopio Nazionale Galileo (TNG),
located at the Roque de los Muchachos Observatory (La Palma,
Canary Islands, Spain). Images were obtained using the Device Optimized
for the LOw RESolution instrument (DOLORES or LRS).
This device has a camera and a spectrograph installed at the Nasmyth
B focus of the telescope. We observed in imaging mode with
the R Johnson filter and a 2$\times$2 binning mode. The camera is equipped with a 2048x2048 CCD with a pixel size of 13.5$\micron$. The field of view is 8.6$\arcmin$$\times$8.6$\arcmin$ with a 0.252$\arcsec$/pix scale (pixel scale for a 1$\times$1 binning). 

In South America, we used the New Technology Telescope (NTT), located at La Silla Observatory (Chile), equipped with the ESO Faint Spectograph and Camera (version 2) or EFOSC2 mounted at the Nasmyth B focus of the telescope. We observed in imaging mode with the R Bessel filter and a 2$\times$2 binning mode. The camera is equipped with a 2048$\times$2048 CCD with a pixel size of 15x15$\micron$ (pixel scale for a 1$\times$1 binning). The field of view is 5.2$\arcmin$$\times$5.2$\arcmin$.

\subsubsection{Regular program on lightcurves of TNOs}

Usually, we studied short-term variability thanks to our program on lightcurves of KBOs at the Sierra Nevada Observatory (OSN) 1.5~m telescope. We present observations carried out at that telescope, at the 2.2~m and the 3.5~m Centro Astronomico Hispano
Aleman (CAHA) telescopes at Calar Alto Observatory (Almeria, Spain) and at the 82~cm telescope of the Instituto de Astrof\'{\i}sica de Canarias (IAC-80 telescope) located at the Teide Observatory (Tenerife, Canary Islands, Spain). 

The OSN observations were carried out by means of a 2k$\times$2k CCD, with a total field of view of 7.8$\arcmin$$\times$7.8$\arcmin$. We used a 2$\times$2 binning mode, which changes the image scale to 0.46$\arcsec$/pixel (pixel scale for a 2$\times$2 binning).

For our CAHA observations, we used the Calar Alto Faint Object Spectrograph (CAFOS) at the 2.2~m telescope and the Large Area Imager for Calar Alto (LAICA) at the 3.5~m telescope. CAFOS is equipped with a 2048$\times$2048 pixel CCD and image scale is 0.53"/pixel (pixel scale for a 1$\times$1 binning). LAICA is equipped with a 2$\times$2 mosaic of 4k$\times$4k CCDs and its total field of view is 44.36$\arcmin$$\times$44.36$\arcmin$ and the pixel scale is 0.225$\arcsec$/pixel. 

We also present here a few observations carried out at the IAC-80 telescope. It is equipped with a 2k$\times$2k CCD camera with a 13.5$\times$13.5~$\micron$/pixel size (pixel scale for a 1$\times$1 binning), installed at the Cassegrain primary focus. Its total field of view is 10.6$\arcmin$$\times$10.6$\arcmin$.

\subsection{Observing strategy}

Exposure times were chosen by considering two main factors. On
one hand, it had to be long enough to achieve a signal-to-noise
ratio (S/N) sufficient to study the observed object (S/N$>$20). On
the other hand, the exposure time had to be short enough to avoid
elongated images of the target (when the telescope was tracked
at sidereal speed) or elongated field stars (if the telescope was
tracked at the TNO rate of motion). We always chose to track the
telescope at sidereal speed. The drift rates of TNOs are typically low, $\sim$2$\arcsec$/h, so exposure times around 300–600 s were typically used.

The OSN, Calar Alto and IAC-80 observations reported in this
work were performed without a filter in order to maximize the S/N. The main goal of our study is short-term variability via relative photometry. Therefore, the use of unfiltered images without absolute calibration is not a problem for our work. The R Bessel and R Johnson filters were used during our observations at the NTT and TNG, respectively. These filters were chosen to maximize the S/N on TNOs while minimizing the fringing that appears at longer wavelengths on images from these instruments.

The targets of our regular programme are typically brighter than
21 mag in V. During our coordinated campaign, we also had the
opportunity to use 4 m class telescopes to observe fainter objects
and select targets with visual magnitudes between 21 and 22.5 mag.
Relevant geometric information about the observed objects on the
dates of observations, the number of images and filters used is
summarized in Table 1

\section{Data reduction and analysis}
\subsection{Data reduction}

During most observing nights, a series of biases and flat-fields were
obtained to correct the instrumental signature from the images. We
thus created a median bias and a median flat-field frame for each
night of observation. Care was taken not to use bias or flat-field
frames that might be affected by observational or acquisition problems.
The median flat-fields were assembled from twilight dithered
images and the results were inspected for possible residuals from
very bright saturated stars. The flat-field exposure times were always
long enough to ensure that no shutter effect was present, so that a
gradient or an artefact of some sort could be present in the corrected images. Each target image was bias subtracted and flat-fielded using
the median bias and median flat-field of the observation night. If
daily information about bias and/or flat-field was not available, we
used the median bias and median flat-field of a former or subsequent
night.

Relative photometry using between 6 and 25 field stars was carried out by means of DAOPHOT routines \citep{Stetson1987}. Care was taken not to introduce spurious results due to faint background stars or galaxies in the aperture. No cosmic ray removal algorithms were used. We rejected images in which the target is affected by a cosmic ray hit or by a nearby star.We used common reduction software for photometry data reduction of all the images adjusting the details of the parameters to the specificity of each data set.

The choice of the aperture radius is important. We had to choose
an aperture as small as possible to obtain the highest S/N by minimizing
the contribution from the sky, but large enough to include
most of the flux of the TNO. Typically,we repeated the measurement
using a set of apertures with radii around the full width at
half-maximum, and also adaptable aperture radius (aperture radius
is varying according to the seeing conditions of each image, and so,
the aperture radius is different for each image). Then, we have to
consider two factors in order to choose the best data reduction: the
aperture size and the reference stars used. For all apertures used, we
chose the results giving the lowest scatter in the photometry of both
targets and stars. Several sets of reference stars were used to establish
the relative photometry of all the targets. In many cases, several
stars had to be rejected from the analysis because they showed some
variability. Finally, the set that gave the lowest scatter was used for
the final result. The final photometry of our targets was computed
by taking the median of all the light curves obtained with respect
to each reference star. By applying this technique, spurious results
were eliminated and the dispersion of photometry was improved.

During the observational campaigns, we tried to stick to the same
field of view, and therefore to the same reference stars, for each
observed target. In some cases, due to the drift of the observed
object, the field changed completely or partially. If the field changed completely, we used different reference stars for two or three subsets
of nights in the entire run. If the field changed partially, we tried
to keep the greatest number of common reference stars during the
whole campaign. In the case of our coordinated campaign using
two telescopes, we tried to observe the same field of view with
both telescopes for any given target. In this way, we can use the
same reference stars and do a better job in image processing and
analysis.
 
When we combined data from several observing runs, we normalized
the photometry data to their average because we did not
have absolute photometry allowing us to link runs. By normalizing
over the averages of several runs, we assume that a similar number
of data points are in the upper and lower part of the curves. This
may not be so if runs were only two or three nights long, which is
not usually the case. We wish to emphasize that we normalized to
the average of each run and not the average of each night

\subsection{Absolute photometry}

We computed approximate R magnitudes for a few images per object per observational run. Namely, we computed approximate magnitudes for the OSN data of 2004~NT$_{33}$ and for all observations of 2002~TC$_{302}$, 2002~TX$_{300}$, 2004~XA$_{192}$, and 2005~UQ$_{513}$. In order to obtain approximate R-magnitudes, we used USNO-B1 stars in the field of view as photometric references. Since the USNO-B1 magnitudes are not standard BVRI magnitudes and since we also did not use BVRI filters, we derived very approximate magnitudes, with typical uncertainties of $\sim$0.4~mag.  
  
During our observations at the TNG and at the NTT, the R Johnson and the R Bessel filters were used, respectively. We are able to report an absolute photometry of all the data carried out during the coordinated campaign. 

For absolute photometry, each image was reduced using standard
techniques of calibration, as presented in this section (bias subtraction
and flat-field correction). During each night of observations,
Landolt standard stars \citep{Landolt1992} were observed at different air masses in order to calculate the calibration parameters such as
photometric zero-point and first-order extinction coefficients.

TNOs are very faint objects, so the choice of the aperture in
order to calculate the flux, or magnitude, is important. In fact, the
aperture must be big enough to collect all the flux of the target
without introducing the contaminating flux of the background. We
computed their fluxes using a small aperture and corrected for the
flux loss by means of the aperture correction \citep{Howell1989, Stetson1990}. We used between 5 and 15 stars in each field of view in
order to compute the aperture correction for each object. The object
aperture radius varied between 3 and 5 pixels, depending on the
brightness of the target and on the night conditions. We chose the aperture that gave the highest S/N for each object by computing the
growth curve of some stars \citep{Howell1989, Stetson1990}. \\

\subsection{Period-detection methods}

The final time-series photometry of each target was inspected for periodicities by means of the Lomb technique \citep{Lomb1976} as implemented in \cite{Press1992}, but we also checked results by means of several other time series analysis techniques, such as Phase Dispersion Minimization (PDM), and CLEAN technique \citep{Foster1995}. \cite{Harris1989} method and its improvement described in \cite{Pravec1996} was also used (hereafter called Pravec-Harris method).

As mentioned before, the reference stars were also inspected for short-term variability and we can thus be confident that no error has been introduced by the choice of reference stars. Finally, in order to measure the amplitudes of short-term variability, we performed Fourier fits to the data to determine the peak-to-peak amplitudes (or full amplitudes).

\section{Photometric results}

In this section, we present our short-term variability results summarized in Table 2. The Lomb periodograms and lightcurves for all objects are provided in Figure 1 to Figure 21. We plotted all lightcurves over two periods. Times for zero phase, without light time correction, are reported in Table~2. For each lightcurves, a Fourier Series is used to fit the photometric data. Error bars for the measurements are not shown on the plots for clarity but one-sigma error bars on the relative magnitudes are reported in the online version of Table 3. Absolute magnitudes are also provided in the online version of Table 3.   

The following subsections are dedicated to the short-term variability of our targets. We organized our results according to the Gladman dynamical classification \citep{Gladman2008}.   

%

\subsection{Classical objects}

%

\subsubsection{(275809) 2001~QY$_{297}$}
The time base ( time coverage between the first and the last image
of the object) of the 2009 data from the NTT is around 10.2 h split
in two nights. The time base of our 2010 data is shorter; 2.3 h
in three nights of observations. For our 2009 data set, the Lomb
periodogram, PDM and CLEAN techniques suggested a rotational
period of around 5.8 h. The Pravec–Harris technique inferred a
double rotational period around 11.6 h. Our 2010 data set is clearly
too short for a period search.

The Lomb periodogram (Fig. 1) of our two data sets showed
three groups of peaks: the first one, with the highest confidence
level, suggested a rotational period of around 5.84 h, the second
one around 4.61 h and the last one around 7.25 h. The CLEAN technique
confirmed a periodic signature at 5.84$\pm$0.34~h. However, PDM presented a single-peaked period of 7.21$\pm$0.39~h, and the Pravec-Harris technique a period around 14.4$\pm$0.6~h and a possible rotational period of 5.84$\pm$0.34~h. 
The best-fitting lightcurve is obtained for a period of 5.84~h (Fig. 2) because the alternative fits show more scatter. The amplitude of the lightcurve is large, 0.49$\pm$0.03~mag assuming a 5.84~h periodicity. 
Assuming that large amplitudes ($>$0.15~mag) are mainly due to shape effects, we must consider the double-peaked lightcurve (see Section 5). Then, if 5.84~h is our preferred photometric period, a preferred rotational period of 11.68~h (2$\times$5.84) is deduced.
    
To our knowledge, a previous study of this target, based on 13 images obtained in around 5~h of observations, was done by \cite{Kern2006} who suggested a rotational period of 12.2$\pm$4.3~h and an amplitude of 0.66$\pm$0.38~mag.   
We conclude, in agreement with \cite{Kern2006}, that 2001~QY$_{297}$ has a moderately long rotational period and a very high amplitude. 

We must point out that 2001~QY$_{297}$ has a satellite. This system is an asynchronous binary system because the primary has a much smaller rotational period than the orbital one. Both components of the system are not resolved in our data, so, we are measuring the magnitude of the pair. The satellite has a long orbital period: 138.11$\pm$0.02~days and it is orbiting at a distance of 9960$\pm$30~km from the primary \citep{Grundy2011}. The magnitude difference between 2001~QY$_{297}$ and its satellite is 0.42$\pm$0.07 \citep{Noll2008}. Due to the orbital and physical characteristics of the system, the satellite contribution to the lightcurve is negligible.     

%

\subsubsection{(307251) 2002~KW$_{14}$}
We observed this target along $\sim$10~h during three nights at the NTT and 0.2~h at the TNG. The Lomb peridogram (Fig. 3) shows a peak with a high confidence level at 4.29~h (5.59~cycles/day) and two aliases, with a lower confidence level, at 5.25~h (4.57~cycles/day) and at 3.69~h (6.49~cycles/day). All the techniques used confirmed a photometric rotational period of 4.29~h or 5.25~h with a similar confidence level. Assuming that the lightcurve is essentially due to the shape of the target, we must consider the double peaked one: the rotational period of 2002~KW$_{14}$ should be 8.58~h or 10.5~h.
Our preferred period is 8.58~h, corresponding to an amplitude of 0.21$\pm$0.03~mag (Fig. 4). However, also a lightcurve fit assuming a rotational period of 10.5~h with an amplitude of 0.26$\pm$0.03~mag is possible.  

%

\subsubsection{(55636) 2002~TX$_{300}$} 
The 2003 data set is already published in \cite{Thirouin2010} in which we concluded that the rotational period of this object should be 8.14$\pm$0.02~h. \cite{Ortiz2004} published a rotational period of 7.89$\pm$0.03 and \cite{Sheppard2003} presented a 8.12~h or a 12.1~h single-peaked rotational lightcurve. 

In Fig. 5, we present the Lomb periodogram of all our dataset (2003, 2009, and 2010) with light-time correction. We note a peak with a high confidence level at 4.08~h (5.89~cycles/day). All used techniques confirmed this period, except Pravec-Harris method which favored a double peaked period at 8.15~h (2.94~cycles/day). A double peaked lightcurve seems to be the best option and it is presented in Fig. 6. However, the possibility of a rotational period around 12~h cannot be excluded. The corresponding amplitude is 0.05$\pm$0.01~mag in all cases.
More data are needed to confirm one of these two possibles rotational periods. Due to the low-amplitude lightcurve of 2002~TX$_{300}$, very high quality of data is needed.        

%

\subsubsection{2004~NT$_{33}$}

We have a time base of $\sim$~16~h at the TNG during three nights and of 19.5~h at the OSN over six nights. The Lomb peridogram (Fig. 7) shows a peak with a high confidence level at 7.87~h (3.05~cycles/day) and two aliases with a lower confidence level at 11.76~h (2.04~cycles/day) and at 5.91~h (4.06~cycles/day). PDM, and CLEAN techniques confirmed the highest peak around 7.8~h. The Pravec-Harris technique suggested a rotational period of 7.87~h, a double-peaked period of 23.52~h, and a possible rotational period of 3.1~h (7.74~cycles/day).
The best-fitting lightcurve is obtained for a period of 7.87~h and a corresponding amplitude of 0.04$\pm$0.01~mag (Fig. 8). 

%

\subsubsection{(230965) 2004~XA$_{192}$}

We have more than 18~h of observation in seven nights in October and more than 7~h during one night in December.
The Lomb periodogram (Fig. 9) shows two peaks with a similar confidence level. The second peak at 7.88~h (3.05~cycles/day) seems to be a little bit higher than the first one at 11.49~h (2.09~cycles/day). PDM and CLEAN techniques confirmed the second peak at 7.88~h, but a period around 11~h is still present with a high confidence level. The Pravec-Harris technique presented a double-peaked period at 15.76~h. In all cases, the amplitude of the curve is 0.07$\pm$0.02~mag.   
A rotational period of 7.88~h appears to be the best option for this object (Fig. 10). The alternative fit of 11.49~h exhibits more scatter and should be probably discarded. 

%

\subsubsection{(202421) 2005~UQ$_{513}$}

The time base of our August run is around 10~h. In September, the time base is 8~h split in three nights and in October, it is around 45~h in six nights   
The Lomb periodogram (Fig. 11) showed one clear peak and a possible 24h-alias. The highest peak is located at 7.03~h (3.41~cycles/day) and the second one is located at 10.01~h (2.40~cycles/day). In Fig. 12, we present both lightcurves. In all cases, the amplitude of the curve is 0.06$\pm$0.02~mag. PDM, CLEAN, Pravec-Harris techniques confirmed these two peaks with a similar spectral power. There is no published photometry for this object, so we cannot compare our results and favor a clearly rotational period.    

%

\subsection{A resonant object}

%

\subsubsection{(84522) 2002~TC$_{302}$}
It is in the 5:2 resonance with Neptune. Time base is, respectively, 10~h over 2 nights, 7~h over two nights and 0.5~h in one night. The Lomb periodogram (Fig. 13) presents three peaks with similar confidence levels. The highest peak is located at 5.41~h (4.44~cycles/day) and two aliases are found at 4.87~h (4.93~cycles/day) and at 6.08~h (3.95~cycles/day). PDM and Pravec-Harris techniques confirmed the highest peak at 5.41~h, but CLEAN favored a rotational period of 6.08~h. The best-fitting lightcurve is obtained for a rotational period of 5.41~h. In Fig. 14, we present the single-peaked lightcurve with an amplitude of 0.04$\pm$0.01 mag. 
%

\subsection{Scattered disc object (SDO) and detached disc objects}

%

\subsubsection{(40314) 1999~KR$_{16}$}

It is a detached disc object. We have less than 30 images for this object, so we cannot present a satisfactory study based only on our data alone. We are just able to estimate an amplitude variation around 0.22~mag in 3.4~h of observations. 

We found data that has been already published about 1999~KR$_{16}$ \footnote{\cite{Rousselot2005} created a database in which lightcurves and photometric data of TNOs can be found (http://www.obs-besancon.fr/bdp/)}. Using their 2001 data set, \cite{Sheppard2002} obtained two best-fit periods of 5.840~h and 5.929~h, but they did not discard the possibility of a double-peaked period.
We merged \cite{Sheppard2002} data and our data in order to obtain an accurate lightcurve. The Lomb periodogram (Fig. 15) shows one peak with a high confidence level, located at 5.80~h (4.14~cycles/day) and two aliases at 7.73~h (3.10~cycles/day) and at 4.73~h (5.08~cycles/day). PDM and CLEAN techniques confirmed the rotational period of 5.8~h. Pravec-Harris method suggested the double-peaked period. In Fig. 16, we present the single-peaked lightcurve. The amplitude of the curve is 0.12$\pm$0.06~mag, which is not at odds with \cite{Sheppard2002}, within uncertainty limits, even if slight differences can be seen, maybe due to the fact that usually \cite{Sheppard2002} perform sinusoidal fits instead of the Fourier series method used in this work. We suggest a rotational period estimation of 5.8~h, close to the one estimated by \cite{Sheppard2002} for this object.
  
%

\subsubsection{(44594) 1999~OX$_{3}$}

It is a scattered disc object. Time base of our data is around 14~h over three nights of observations. The Lomb periodogram (Fig. 17) shows several peaks. The highest one is found at 15.45~h (1.55~cycles/day). We note two aliases at 9.26~h (2.59~cycles/day) and at 36.92~h (0.65~cycles/day). The PDM technique favored the peak around 9~h. CLEAN shows two peaks with a similar confidence level around 9~h and 15~h.
The Pravec-Harris method favored three possible rotational periods: 9.26~h, 13.4~h, and 15.45~h. In Fig. 18 and Fig. 19, we present all lightcurves. The amplitude of the curves is  0.11$\pm$0.02~mag. To our knowledge, there is no published photometry for this object to compare with.    

%

\subsubsection{(145480) 2005~TB$_{190}$}

It is a detached disc object. The Lomb periodogram (Fig. 20) shows one peak with a high confidence level and two aliases with a lower confidence level. 
The highest peak is located at 12.68~h (1.89~cycles/day) and the two aliases are located at 28.57~h (0.84~cycles/day) and at 8.16~h (2.94~cycles/day). All techniques confirm a rotational period of 12.68~h for this target, as shown in Fig. 21 for the single-peaked lightcurve. The Pravec-Harris technique favored two possible rotational periods: 12.68~h and 16.32~h (2$\times$8.16~h).
Our first estimation of 12.68~h seems to be the best option. The amplitude of the curve is 0.12$\pm$0.01~mag. 
 
%
\section{Discussion}

The detached disc object, 2005~TB$_{190}$, is a paradigmatic example of the efficiency of having coordinated campaigns. In fact, during the first two nights, we managed to coordinate observations from the Canary Islands and Chile, observing this body on the first night during 2.2~h at the TNG, and around 4~h at the NTT, allowing us to study close to a half-period on one single coordinated night. Finally, with less than 50 images in 4 nights, we could reliably estimate the moderately long rotational period for this object. Detection and reliable estimations of long rotational periods were one of the goals of this coordinated campaign. We would have probably needed many more images and detection of this long periodicity would have probably been difficult without a coordinated campaign. Thus we consider our first coordinated campaign as a successful beginning.

As a general feature of our results, we report that the average amplitude of our sample is 0.13~mag. We note that in our 10 objects sample, only 2 have amplitudes larger than 0.15~mag.
\cite{Thirouin2010} and \cite{Duffard2009} suggested a threshold of 0.15~mag in order to distinguish among lightcurve variations due to albedo or due to the shape of the target because the best fits to Maxwellian distributions were obtained with that assumption. Low amplitudes can be explained by albedo heterogeneity on the surface of a MacLaurin spheroid, while large amplitudes of variability are probably due to the shape of an elongated Jacobi body. According to this assumption, we introduce the criterion to consider that a high lightcurve amplitude of a large object may be attributed to a non spherical shape (typically a triaxial ellipsoid). In this case, we prefer the double peaked lightcurve to represent a complete rotation of the object. We must point out that to distinguish between shape and albedo contribution in a lightcurve is not trivial at all. 
In Table 2, we indicate the rotational periods obtained from data reduction (preferred photometric period) and the preferred rotational period assuming our criterion. For example, in the case of 2001~QY$_{297}$, our data analysis suggests a rotational period of 5.84~h, but given an amplitude larger than 0.15~mag, the amplitude variation is probably due to the shape of the object and we prefer the double peaked period, 11.68~h (2$\times$5.84~h) as true rotational period of the object. 

In Fig. 22, we plot the lightcurve peak-to-peak amplitude versus the absolute magnitude of results shown in this work and already published in \cite{Thirouin2010}. As shown in Fig. 22, the majority of studied objects present a low amplitude, typically $<$~0.15~mag. In fact, except some cases like 2001~QY$_{297}$, most TNOs have a low amplitude. We found an average amplitude of 0.09~mag, 0.11~mag, 0.12~mag and 0.10~mag for, respectively, the scattered/detached, the resonant, the classical and the centaur groups. So, there is not a dynamical group with a higher/smaller amplitude in our database.        
We must point out that the lack of long rotational periods, previously mentioned in \cite{Thirouin2010} seems to be confirmed by this new work. In fact, except 2005~TB$_{190}$ and probably 2001~QY$_{297}$, all our targets present a rotational period $<$10~h.    

Assuming TNOs in general as triaxial ellipsoids, with axes a$>$b$>$c (rotating along c), the lightcurve amplitude, $\Delta{m}$, varies as a function of the observational angle $\xi$ (the angle between the rotation axis and the line of sight) according to \cite{Binzel1989}:
\begin{equation}
\Delta{m} = 2.5~log\left( \frac{a}{b}\right)  - 1.25~log\left(
\frac{a^{2}\cos^{2}\xi + c^{2}\sin^{2}\xi}{b^{2}\cos^{2}\xi +
c^{2}\sin^{2}\xi}\right)
\end{equation}
We computed a lower limit for the object elongation (a/b), assuming an equatorial view ($\xi$ = 90$^{\circ}$)
\begin{equation}
\Delta{m} = 2.5~log\left(\frac{a}{b}\right)
\end{equation}
According to \cite{Chandrasekhar1987} study of figures of equilibrium for fluid bodies, we can estimate lower limits for densities from rotational periods and the elongation of objects. That is to say, assuming that a given TNO is a triaxial ellipsoid in hydrostatic equilibrium (a Jacobi ellipsoid), we can compute a lower density limit. This study is summarized in Fig. 23 which is an update of fig. 7 of \cite{Duffard2009}. 
In our sample, only two bodies have a high amplitude lightcurve ($>$0.15~mag) and can be assumed to be Jacobi ellipsoids: 2001~QY$_{297}$ and 2002~KW$_{14}$. 2001~QY$_{297}$ has a very low density if it is in hydrostatic equilibrium and 2002~KW$_{14}$ seems to have a density between 0.5 and 1~g cm$^{-3}$.
Using Equation 1, we compute the lower limit for the densities of these two bodies, assuming a viewing angle of 60$^{\circ}$ \footnote{Given a random distribution of spin vectors, the average of viewing angle is 60$^{\circ}$.}. The results are reported in Table 2. 
Most of our targets have low-amplitude lightcurves, probably due to albedo effects. So, they are probably MacLaurin spheroids and the study on lower limit densities cannot be applied. In fact, most of observed objects are far from the theoretical curves for acceptable values for the density which indicates that those objects are likely MacLaurin spheroids or are not in hydrostatic equilibrium (Fig. 23).
\\
TNOs densities are an important physical characteristic. Unfortunately, their estimation is complicated and usually obtained just for binary and multiple systems. The range of published densities varies from around 1~g cm$^{-3}$ for Varuna \citep{Jewitt2002} to 4.2$\pm$1.3~g~cm$^{-3}$ for Quaoar \citep{Fraser2010} (however, a recent stellar occultation by Quaoar indicates that Quaoar density is probably much smaller than published one \citep{Braga-Ribas2011}). Generally, densities are supposed to be very low in the Edgeworth-Kuiper Belt $\aplt$1~g cm$^{-3}$, except for some ``atypical" cases such as Haumea, Eris, and Pluto. 

We also studied an asynchronous binary classical belt object, 2001~QY$_{297}$. 
We find a large lightcurve amplitude for this object, (0.49$\pm$0.03)~mag. A rotational period of 11.68~h seems to be the best candidate. The lightcurve of this object is likely due to its shape. Assuming that 2001~QY$_{297}$ is a triaxial ellipsoid in hydrostatic equilibrium, we estimate large axis ratios: b/a around 0.64 and c/a around 0.45.
\\
If we assume that 2001~QY$_{297}$ is in hydrostatic equilibrium, we can estimate its bulk density, $\rho$, according to \cite{Chandrasekhar1987}, and define the volume of the system as V$_{sys}$=M$_{sys}$$/$$\rho$. Assuming that its rotational period is 11.68~h, we estimate a lower limit density of $\rho$=290~kg$~$m$^{-3}$. 
Assuming that both components have the same albedo, we work out the primary radius by
\begin{equation}
R_{primary} = \left( \frac{3V_{sys}}{4\pi\left(1+10^{-0.6\Delta_{mag}}\right)}\right)^{1/3}
\end{equation}
where R$_{primary}$ is the radius of the primary and $\Delta_{mag}$ is the component magnitude difference.  
Assuming that both components have the same albedo, we expressed the satellite radius as: 
\begin{equation}
R_{satellite} = R_{primary}10^{-0.2\Delta_{mag}}
\end{equation}
with a $\Delta_{mag}$=0.42 \citep{Noll2008} and a density, $\rho$=290~kg$~$m$^{-3}$, we computed a primary radius of 129~km and a satellite radius of 107~km for a total mass of the system, M$_{sys}$=(4.105$\pm$0.038)x10$^{18}$~kg \citep{Grundy2011}.   
The effective radius of the system is expressed as:
\begin{equation}
R_{effective} = \sqrt{R_{primary}^{2}+R_{satellite}^{2}}
\end{equation}
By using primary and secondary sizes obtained before, we computed an effective radius of 168~km for this system.
We can derive the geometric albedo, p$_{\lambda}$, that is given by the equation:
\begin{equation}
p_{\lambda} =  \left(\frac{C_{\lambda}}{R_{effective}}\right)^{2}10^{-0.4H_{\lambda}}
\end{equation}
C$_{\lambda}$ is a constant depending on the wavelength \citep{Harris1998}, and H is  the absolute magnitude. The value we find for the geometric albedo is 0.08. 
\\
Assuming spherical shapes and densities between 500 and 2000~kg$~$m$^{-3}$, \cite{Grundy2011} published an albedo range of 0.13-0.32. They also reported a primary radius ranging from 64 to 100~km (values obtained assuming spherical shapes and densities between 500 and 2000~kg$~$m$^{-3}$).
Due to the facts that 2001~QY$_{297}$ has a low inclination (1.5$^{\circ}$, according to the Minor Planet Center (MPC) database) and a high albedo, \cite{Grundy2011} concluded that this body belongs to a more excited class of small TNOs \citep{Brucker2009}.
\\
According to our study, 2001~QY$_{297}$ has instead a low albedo. Both studies, (\cite{Grundy2011} and our estimation) are preliminary, but \textit{Herschel Space Observatory} key program ``TNO's are Cool!" estimated the albedo and the size of this binary object \citep{Vilenius2012}. 

Various models can be enumerated in order to explain the formation of binary or multiple systems. Models based on gravitational capture have already been presented \citep{Goldreich2002, Astakhov2005}, as well as models based on low velocity collision between Kuiper Belt Objects \citep{Durda2004} or the gravitational collapse model \citep{Nesvorny2010}. Recently, the possibility of rotational fission in the Kuiper Belt has been considered in \cite{Ortiz2012}.
\\
We computed the specific angular momentum of 2001~QY$_{297}$ system using the formula published in \cite{Descamps2008} and the scaled spin rate according to \cite{Chandrasekhar1987}. The specific angular momentum of this binary is 1.61$\pm$0.13 and its scaled spin rate is 0.61$\pm$0.01 (specific angular momentum and scaled spin rate are adimensional values). Those values seem to indicate that the 2001~QY$_{297}$ binary system was not formed by rotational fission. In fact, the high value of the specific angular momentum and the scaled spin rate of this system do not fall into the ``high size ratio binaries" region indicated in the fig. 1 of \cite{Descamps2008}. So, we can probably discard a possible rotational fission origin for this binary. We cannot favor any other formation scenario; this system could have been formed by capture and/or collision, or gravitational collapse.    

The last part of this section is dedicated to two examples of solar phase curves. The phase function can be expressed in flux as
\begin{equation}
\phi(\alpha) = 10^{-0.4\beta \alpha}
\end{equation}
where $\alpha$ is the phase angle (in degrees) and $\beta$ is the phase coefficient in magnitudes per degree at phase angles $<$2$^\circ$.  
All our targets were observed in a range of phase angles insufficient to perform a reliable study of the solar phase curve. Using various datasets already published, we report solar phase curves of (40314) 1999~KR$_{16}$ and of (44594) 1999~OX$_{3}$. Distance correction was applied and brightness variations due to rotation were removed to R-band magnitudes (R-band absolute magnitudes of TNG and NTT data in the online version of Table 3). Corrected R-band magnitudes will be called  m$_{R}$(1,1,$\alpha$) hereinafter, indicating with $\alpha$ the phase angle, ``1" stands for 1 AU (geocentric and heliocentric distances). For observations done at the same phase angles, we averaged magnitudes and computed corresponding uncertainties. 

In Fig. 24, we plot the solar phase curve of (40314) 1999~KR$_{16}$. According to \cite{Sheppard2002} and to data reported in this work, we obtain a phase angle range of around 1.5$^{\circ}$,  m$_{R}$(1,1,$\alpha$)~=~5.41$\pm$0.03~mag and $\beta$~=~0.12$\pm$0.03~mag$\cdot$deg$^{-1}$. These results are consistent with \cite{Sheppard2002}, who found m$_{R}$(1,1,$\alpha$)~=~5.37$\pm$0.02~mag and $\beta$~=~0.14$\pm$0.02~mag$\cdot$deg$^{-1}$. 

Fig. 25 shows the solar phase curve of (44594) 1999~OX$_{3}$, based on \cite{Bauer2003} and on our data. We get m$_{R}$(1,1,$\alpha$)~=~6.65$\pm$0.03~mag and $\beta$~=~0.30$\pm$0.03~mag$\cdot$deg$^{-1}$ from all data. \cite{Bauer2003} reported  m$_{R}$(1,1,$\alpha$)~=~7.1~mag, uncorrected for phase angle and for possible rotation. 
Assuming  albedo values of 0.25 and 0.05, we derived the conversion from m$_{R}$(1,1,$\alpha$) to size, obtaining, respectively, size estimations of 130 and 300 km for 1999~OX$_{3}$.
Assuming the same albedo values, we finally obtained a size range of 210-470~km for 1999~KR$_{16}$.

%

\section{Conclusions}

We have collected and analyzed R-band and Clear-band photometric data for TNOs in order to increase the number of objects studied so far. We have reported our first coordinated campaign for TNOs. Coordinating two telescopes, one in Chile and one in the Canary Islands allowed us to monitor during a long time our targets and to try to minimize aliases in the data analysis. We also report our latest result on short-term variability from our regular program of TNOs. We present a homogeneous dataset composed of 10 TNOs. Two of 10 objects (20 per cent) in our sample (2001~QY$_{297}$ and 2002~KW$_{14}$) show a lightcurve with an amplitude $\Delta$$_{m}$~$\geq$0.15~mag. In an extended sample combining objects from this work and from \cite{Thirouin2010}, we computed that 8 of 37 (22 per cent) targets have a $\Delta$$_{m}$~$\geq$0.15~mag. Two of 10 objects (20 per cent) in our sample (2001~QY$_{297}$ and 2005~TB$_{190}$) have a rotational period P$_{rot}$~$\geq$10~h. In an extended sample combining objects from this work and from \cite{Thirouin2010}, we computed that 5 of 37 (14 per cent) targets have a P$_{rot}$~$\geq$10~h. In fact, the sample of studied targets, in the literature, is highly biased toward objects with a short rotational period. The best option to debias the sample, and study objects with a medium to long rotational periodicity, is to carry out coordinated campaigns with two or three telescopes around the world.     

In our sample, 80 per cent of the studied objects have a low variability (less than 0.15~mag) and corresponding lightcurves could be explained by albedo variations. Such bodies are probably MacLaurin spheroids. Just two of 10 objects (2001~QY$_{297}$ and 2002~KW$_{14}$) can be considered Jacobi ellipsoid with a high amplitude lightcurve, probably due to the shape of the body.  
 
We also have studied a binary KBO which turned out to be asynchronous: 2001~QY$_{297}$ which presents a very high variability ($>$~0.4~mag) and a rotational periodicity longer than 10~h. Assuming that the system is in hydrostatic equilibrium and has a very low density, we derived a primary radius of 129~km, a secondary radius of 107~km and a geometric albedo of 0.08 for both components. We examined several possible formation scenarios. This binary was not likely formed by rotational fission due to its high specific angular momentum. We favor a collisional and/or capture scenario, however, a formation based on gravitational instability cannot be ruled out. 

\section*{Acknowledgments}

We are grateful to the Sierra Nevada Observatory, Calar Alto, Telescopio Nazionale Galileo, New Technology Telescope, and El Teide Observatory staffs. This research was based on data obtained at the Observatorio de Sierra Nevada which is operated by the Instituto de Astrof\'{i}sica de Andaluc\'{i}a, CSIC. This research is also based on observations collected at the Centro Astron\'{o}mico Hispano Alem\'{a}n (CAHA) at Calar Alto, operated jointly by the
Max-Planck Institut f\"{u}r Astronomie and the Instituto de Astrof\'{i}sica de Andaluc\'{i}a (CSIC). Other results were obtained at the Telescopio Nazionale Galileo. The Telescopio Nazionale Galileo (TNG) is operated by the Fundaci\'{o}n Galileo Galilei of the Italian Istituto Nazionale di Astrofisica (INAF) on the island of La Palma in the Spanish Observatorio del Roque de los Muchachos of the Instituto de Astrof\'{i}sica de Canarias. Some results are based on observations made with ESO New Technology Telescope (NTT) at the La Silla Observatory under programme ID 083.C-0642A. Some of the data published in this article were acquired with the
IAC-80 telescope operated by the Instituto de Astrof\'{i}sica de Canarias in the Observatorio del Teide. AT, JLO, NM, and RD were supported by AYA2008-06202-C03-01, and ACB was supported by AYA2008-06202-C03-03, which are Spanish MICINN and MEC projects. AT, JLO, NM, and RD also acknowledge the Proyecto de Excelencia de la Junta de Andaluc\'{i}a, J.A.2007-FQM2998. RD acknowledges financial support from the MICINN (Ramon y Cajal fellowship). FEDER funds are also acknowledged.

\begin{onecolumn}
\begin{longtable}{lccccccc}
\caption{Dates (UT-dates), heliocentric (r$_{h}$), and geocentric ($\Delta$) distances and phase angle ($\alpha$) of the observations. We also indicate the number of images used for this work and the number of taken images.  For example, 1/5 indicates that 5 images were taken during our run but just 1 was used for this work. We also summarized the filter used and the telescope for each observational run.} \\
Object  & Date & \# Images & r$_{h}$ [AU] &
$\Delta$ [AU] & $\alpha$[deg] & Filter & Telescope \\
\hline
1999~KR$_{16}$ 
          &  26/07/2009  & 12/16  & 36.034 & 35.913  & 1.61  &  R & NTT \\
          &  27/07/2009  & 7/12 & 36.034  & 35.929 &  1.61  & R  & NTT\\    
1999~OX$_{3}$ 
          &  25/07/2009  & 16/18  & 22.433 & 21.545 &1.29  & R  & NTT  \\
          &  26/07/2009  & 11/23 & 22.431 & 21.536  & 1.25 & R & NTT  \\    
          &  27/07/2009  & 15/19 & 22.430 & 21.527  & 1.21  & R & NTT \\     
2001~QY$_{297}$   
          &  24/07/2009 & 2/5  &   43.142 & 42.168  & 0.39 &  R &  TNG\\
          &  24/07/2009 &16/22 &   43.142  & 42.168  & 0.38 &  R &  NTT\\
          &  25/07/2009 &10/10 &  43.143& 42.166   & 0.36 & R  & NTT  \\    
          &  05/08/2010 &10/10 &  43.223 & 42.215  & 0.15 &  R &  NTT\\
          &  13/08/2010 &6/7 &  43.225& 42.212  & 0.04 & R  & NTT  \\    
          &  14/08/2010 &6/6 & 43.225  & 42.213 & 0.06 &  R &  NTT\\
2002~KW$_{14}$
          &  24/07/2009  & 3/3 & 40.655 & 40.149 & 1.25 & R  & TNG \\
          &  25/07/2009  & 11/16 & 40.656  & 40.167 & 1.26 & R & NTT \\
          &  26/07/2009 & 17/18  & 40.656 &  40.182 & 1.27 & R & NTT \\    
          &  27/07/2009  & 14/16  & 40.657  & 40.197 & 1.28 & R & NTT \\    
2002~TC$_{302}$
          & 15/10/2009   & 13/15  & 46.552 & 45.589  & 0.32 &  Clear & 2.2~m Calar Alto telescope  \\
          &  17/10/2009  & 19/21 & 46.551 & 45.582 & 0.28 &  Clear & 2.2~m Calar Alto telescope  \\
          &  09/09/2010  & 22/23 & 46.331 & 45.684 & 0.96 &  Clear & OSN  \\
          &  11/09/2010  & 10/11 & 46.329 & 45.656 &  0.93 &  Clear & OSN   \\
          &  01/12/2010  & 6/6  & 46.275 &  45.463 &  0.70 &  Clear & IAC-80   \\
2002~TX$_{300}$
          &  07/08/2003  & 116/127 & 40.825 & 40.303 & 1.23 & Clear & OSN \\
          &  08/08/2003  & 165/177 & 40.825 & 40.291 & 1.22 & Clear & OSN \\
          &  09/08/2003  & 132/173 & 40.825 & 40.278 & 1.20 & Clear & OSN \\  
          & 18/10/2009   & 14/19  & 41.534 & 40.615   & 0.54 & Clear & 2.2~m Calar Alto telescope   \\ 
          & 06/09/2010   & 9/14  & 41.639 & 40.901   & 0.95 & Clear & OSN  \\ 
          & 07/09/2010   & 4/7   & 41.639 & 40.891   & 0.94 & Clear & OSN  \\ 
          & 08/09/2010   & 25/25 & 41.639 & 40.884   & 0.92 & Clear & OSN  \\ 
          & 09/09/2010   & 13/19  & 41.640 & 40.875   & 0.91 & Clear & OSN  \\ 
          & 10/09/2010   & 34/36  & 41.640 & 40.867   & 0.90 & Clear & OSN  \\ 
          & 11/09/2010   & 5/5  & 41.640 & 40.857   & 0.88 & Clear & OSN  \\ 
2004~NT$_{33}$  
          & 25/07/2009   & 14/14  & 38.164 & 37.327 & 0.87 & R  & TNG \\
          & 26/07/2009  & 11/11   & 38.164 & 37.234 & 0.87 &  R &  TNG \\    
          & 27/07/2009   & 11/21 &  38.164 & 37.321 & 0.86 &  R  & TNG \\     
          & 13/10/2009   & 15/15  & 38.185  & 37.783  & 1.38 & Clear & OSN  \\
          & 14/10/2009   & 19/20  & 38.185 & 37.796 & 1.39 & Clear & OSN  \\
          & 15/10/2009   & 15/15  & 38.185 & 37.810 & 1.39 & Clear & OSN  \\
          & 16/10/2009   & 12/15  & 38.186 & 37.824 & 1.40 & Clear & OSN  \\
          & 17/10/2009   & 15/20  & 38.186 & 37.837 & 1.41 & Clear & OSN  \\
          & 18/10/2009   & 10/20  & 38.186 & 37.851 & 1.41 & Clear & OSN  \\
2004~XA$_{192}$ 
          & 13/10/2009   & 12/12  & 35.799 & 35.507 & 1.53 & Clear & OSN  \\
          & 14/10/2009   & 6/10  & 35.799 & 35.494 & 1.52 & Clear & OSN  \\
          & 15/10/2009   & 10/10  & 35.799 & 35.481 & 1.52 & Clear & OSN  \\
          & 16/10/2009   & 10/10  & 35.799 & 35.467 & 1.51 & Clear & OSN  \\
          & 17/10/2009   & 22/24  & 35.799 & 35.454 & 1.50 & Clear & OSN  \\
          & 18/10/2009   & 13/13  & 35.798  & 35.439 & 1.49 & Clear & OSN  \\
          & 17/12/2009   & 31/33  & 35.787  & 34.978 & 0.91 & R & 3.5~m Calar Alto telescope \\
2005~TB$_{190}$ 
          &   24/07/2009     & 6/6 & 46.396 & 45.650  &  0.86 &R  &  TNG\\
          &    24/07/2009    & 10/24 &46.396  & 45.650 & 0.86  & R  & NTT \\    
          &   25/07/2009     & 7/11 &46.396  & 45.638 & 0.84 & R &  TNG\\     
          &    25/07/2009    & 5/8 & 46.396 & 45.638  & 0.84  & R & NTT \\
          &   26/07/2009     & 8/8 & 46.396 &  45.627 &0.82  & R & TNG \\     
          &    27/07/2009    & 12/12 & 46.396 &  45.616 & 0.81 & R & TNG \\
2005~UQ$_{513}$
          &  02/08/2008  & 5/10 & 48.806 & 48.389 & 1.09  & Clear & OSN \\
          &  03/08/2008  & 7/13 & 48.806  & 48.376 & 1.08 & Clear & OSN \\
          &  04/08/2008  & 13/15 &  48.806 & 48.362 & 1.07 & Clear & OSN \\
          &  09/08/2008  & 20/25 &  48.805 & 48.294 & 1.03 & Clear & OSN \\
          &  20/09/2009  & 15/18  &48.735  & 47.859 & 0.58 & Clear & OSN \\
          &  21/09/2009  & 38/41 & 48.735 & 47.855 & 0.57 & Clear & OSN \\
         &  23/09/2009  & 18/19 & 48.735 & 47.847 & 0.55 & Clear & OSN \\
          &  13/10/2009  & 33/35 & 48.731  & 47.826 & 0.50 & Clear & OSN \\
          &  14/10/2009  & 30/35 & 48.731  & 47.828 & 0.50 & Clear & OSN \\
          &  15/10/2009  & 30/30 &  48.731 & 47.830 & 0.51 & Clear & OSN \\
          &  16/10/2009  & 24/25 & 48.731  & 47.832 &  0.51& Clear & OSN \\
          &  17/10/2009  & 31/35 & 48.731  & 47.834 &  0.52& Clear & OSN \\
          &  18/10/2009  & 10/14 & 48.731  & 47.837 & 0.52 & Clear & OSN \\
\hline\hline
\end{longtable}
\end{onecolumn}

\begin{onecolumn}
\begin{scriptsize}
\begin{table*}
\caption{\label{tab3}Summary of results from this work. In this table, we present the name of the object, the preferred period (Pref. rot. per. in hour), the preferred photometric period (Pref. phot. per. in hour) and lightcurve amplitude (Amp. in magnitude), the Julian Date ($\varphi_{0}$) for which the phase is zero in our lightcurves (without light time correction), and the absolute magnitudes (Abs. mag.) (Absolute magnitudes extracted from the MPC database). Lower limit to the densities are also shown for two objects (see text). The preferred photometric period is the periodicity obtained thanks to the data reduction. In some cases, as mentioned in the Photometric Results and Discussion sections, we preferred the double rotational periodicity due to the high amplitude lightcurve (the preferred period). 
Zero phase of (40314) 1999~KR$_{16}$ extracted from Sheppard $\&$ Jewitt (2002). }
\begin{tabular}{lcccccc}
\hline
 Object  & Pref. phot. per. [h] & Pref. rot. per. [h]& Amp. [mag.] & $\varphi_{0}$ [JD]& Abs. mag. & $\rho$[g/cm$^{3}$] \\
\hline
(40314) 1999~KR$_{16}$  & 5.8 & 5.8 &  0.12$\pm$0.06  & 2451662.9409  & 5.8 &     \\
(44594) 1999~OX$_{3}$   & 9.26 or 13.4 or 15.45 & 9.26 or 13.4 or 15.45 &  0.11$\pm$0.02  & 2455038.69404  & 7.4 &   \\
(275809) 2001~QY$_{297}$  & 5.84 &  11.68  & 0.49$\pm$0.03  & 2455037.61147  & 5.7 & 0.29 \\
(307251) 2002~KW$_{14}$  & 4.29 or 5.25 &  8.58 or 10.5 & (0.21 or 0.26)$\pm$0.03 &2455037.40786 & 5.0 &  0.53 or 0.35   \\
(84522) 2002~TC$_{302}$ &   5.41  & 5.41 &  0.04$\pm$0.01  & 2455120.41362  & 3.8 &    \\
 (55636) 2002~TX$_{300}$ &   8.15 or 11.7  & 8.15 or 11.7 &  0.05$\pm$0.01  &2452859.51500 & 3.3 &    \\
2004~NT$_{33}$   &  7.87 & 7.87  &  0.04$\pm$0.01  & 2455038.48984  & 4.4 &  \\
(230965) 2004~XA$_{192}$ &  7.88  & 7.88  &  0.07$\pm$0.02  & 2455118.50584  & 4.0 &   \\
(145480) 2005~TB$_{190}$  &  12.68 & 12.68 &  0.12$\pm$0.01  &2455037.62904 & 4.7 &     \\
(202421) 2005~UQ$_{513}$  &   7.03 or 10.01  & 7.03 or 10.01 &  0.06$\pm$0.02  &2455118.32179 & 3.4 &    \\
\hline\hline
\end{tabular}
\end{table*}
\end{scriptsize}
\end{onecolumn}

\clearpage

\begin{figure*}
\begin{center}
\includegraphics[width=8cm, angle=90]{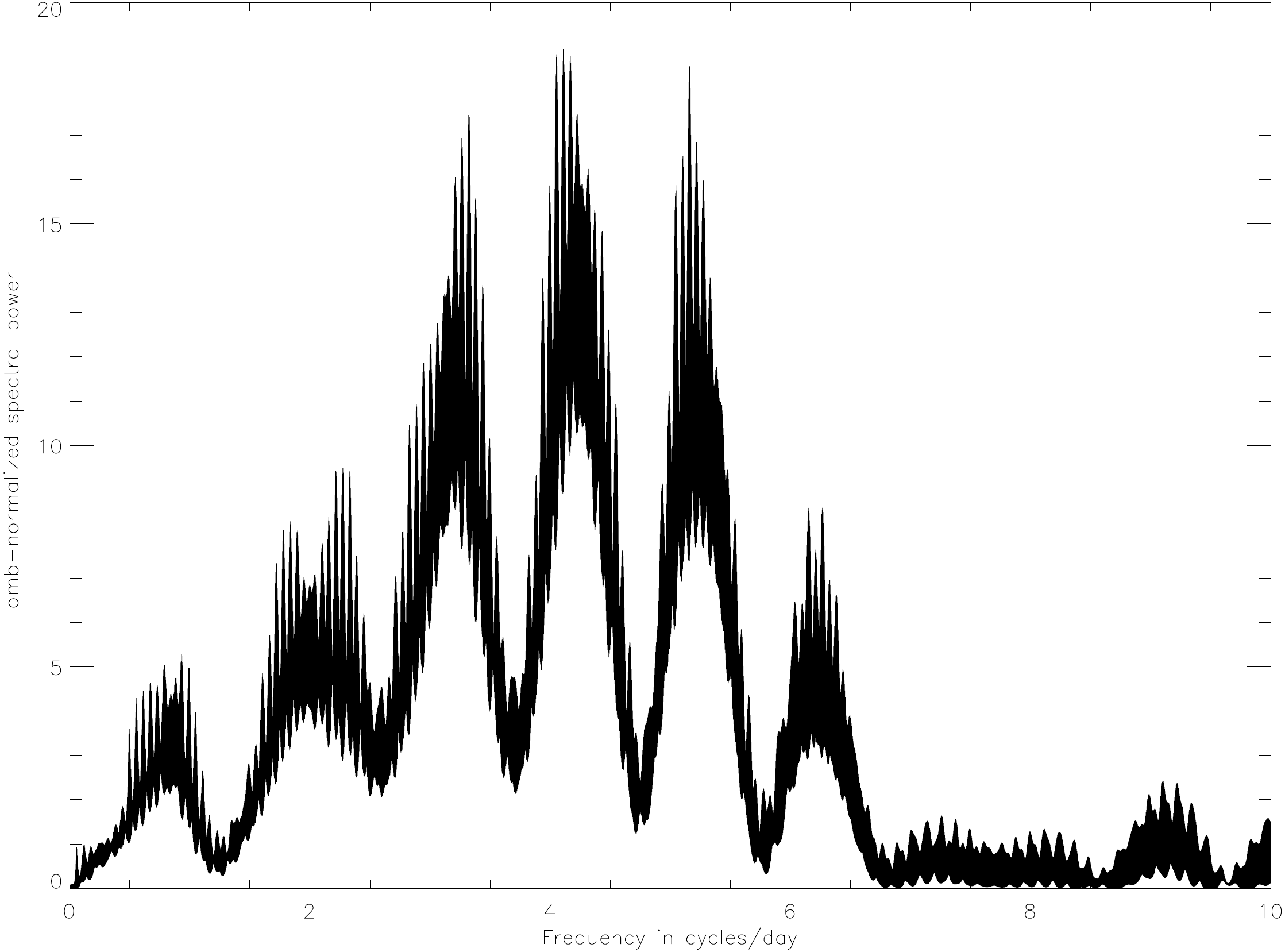}
\caption{Lomb periodogram of 2001~QY$_{297}$}
\end{center}
\end{figure*}

\clearpage

\begin{figure*}
\begin{center}
\includegraphics[width=8cm, angle=90]{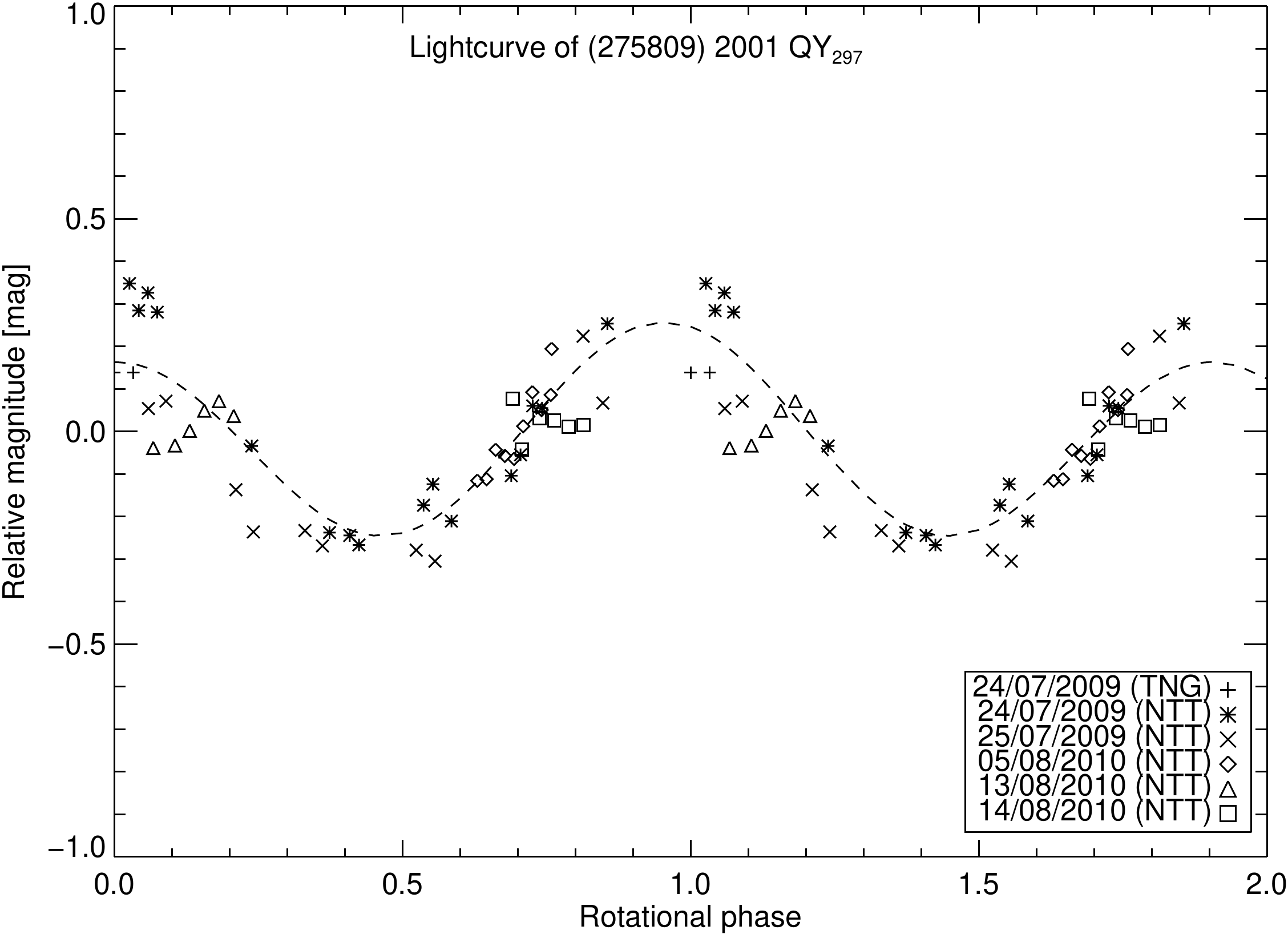}
\includegraphics[width=8cm, angle=90]{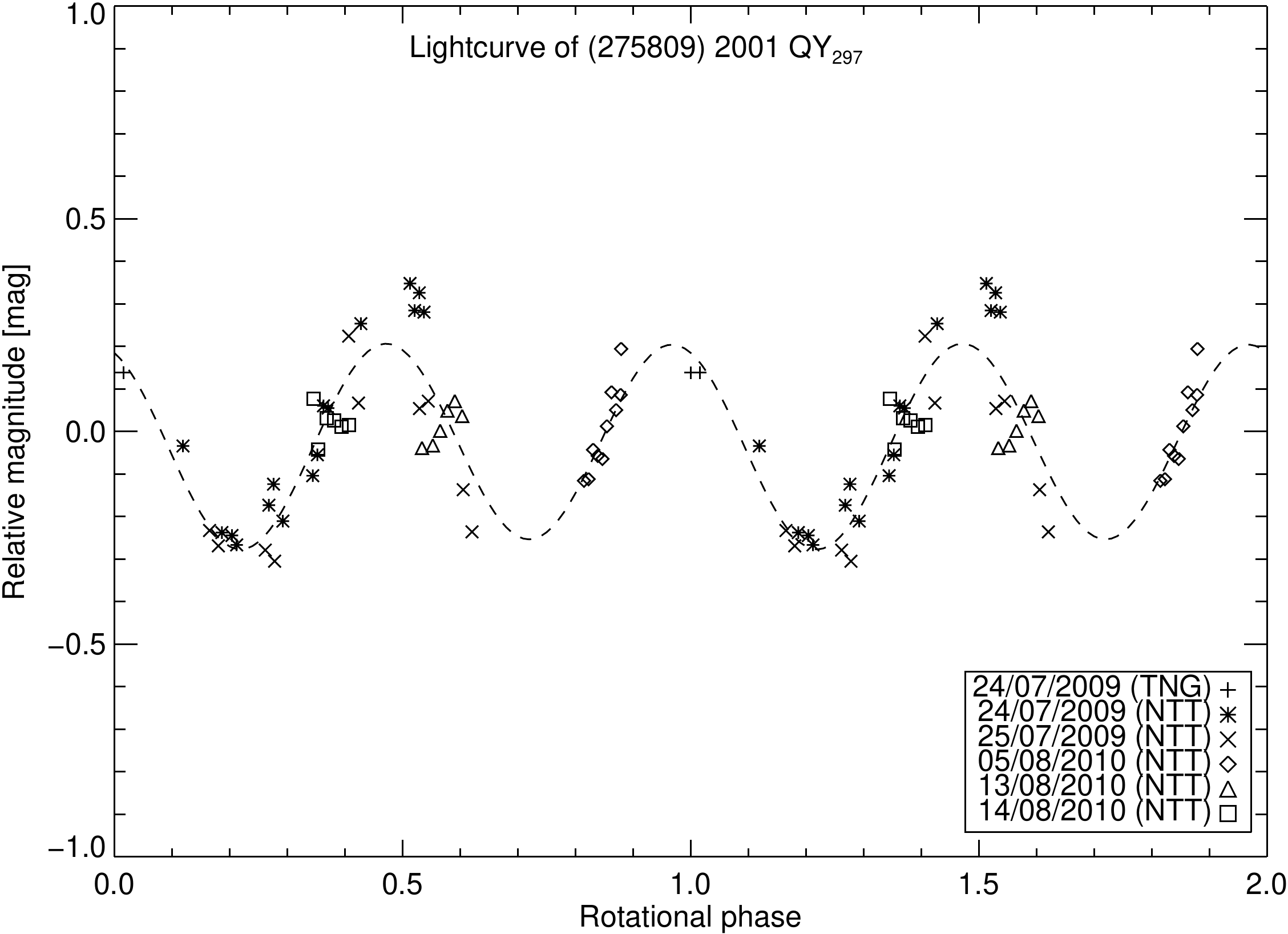}
\caption{Rotational phase curves for 2001~QY$_{297}$ obtained by using a spin
period of 5.84~h (upper plot) and 11.68~h (lower plot). The dash line is a Fourier Series fit of the photometric
data. Different symbols correspond to different dates.}
\label{fig2}
\end{center}
\end{figure*}

\clearpage

\begin{figure*}
\begin{center}
\includegraphics[width=8cm, angle=90]{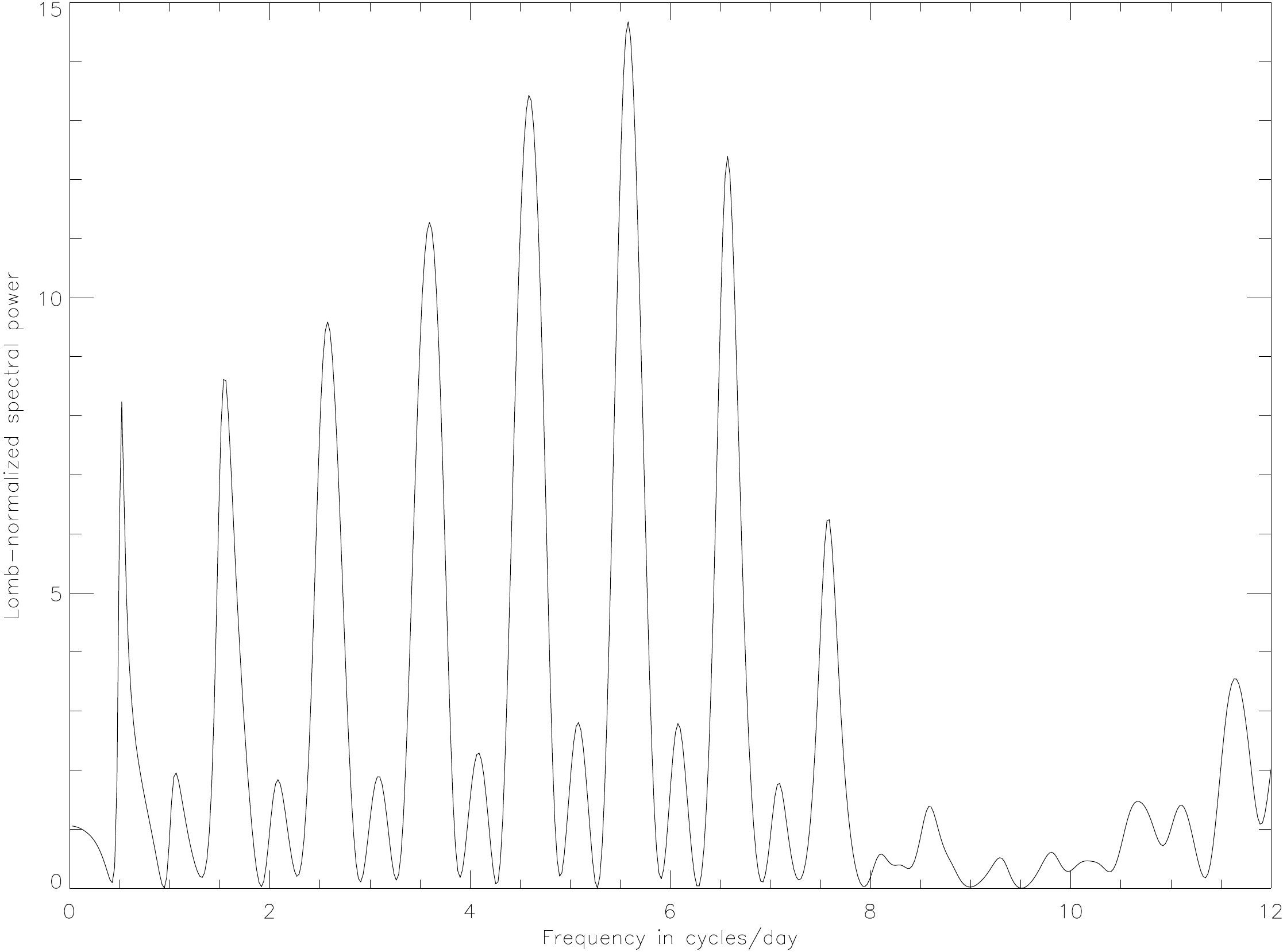}
\caption{Lomb periodogram of 2002~KW$_{14}$}
\label{fig5}
\end{center}
\end{figure*}

\clearpage

\begin{figure*}
\begin{center}
\includegraphics[width=8cm, angle=90]{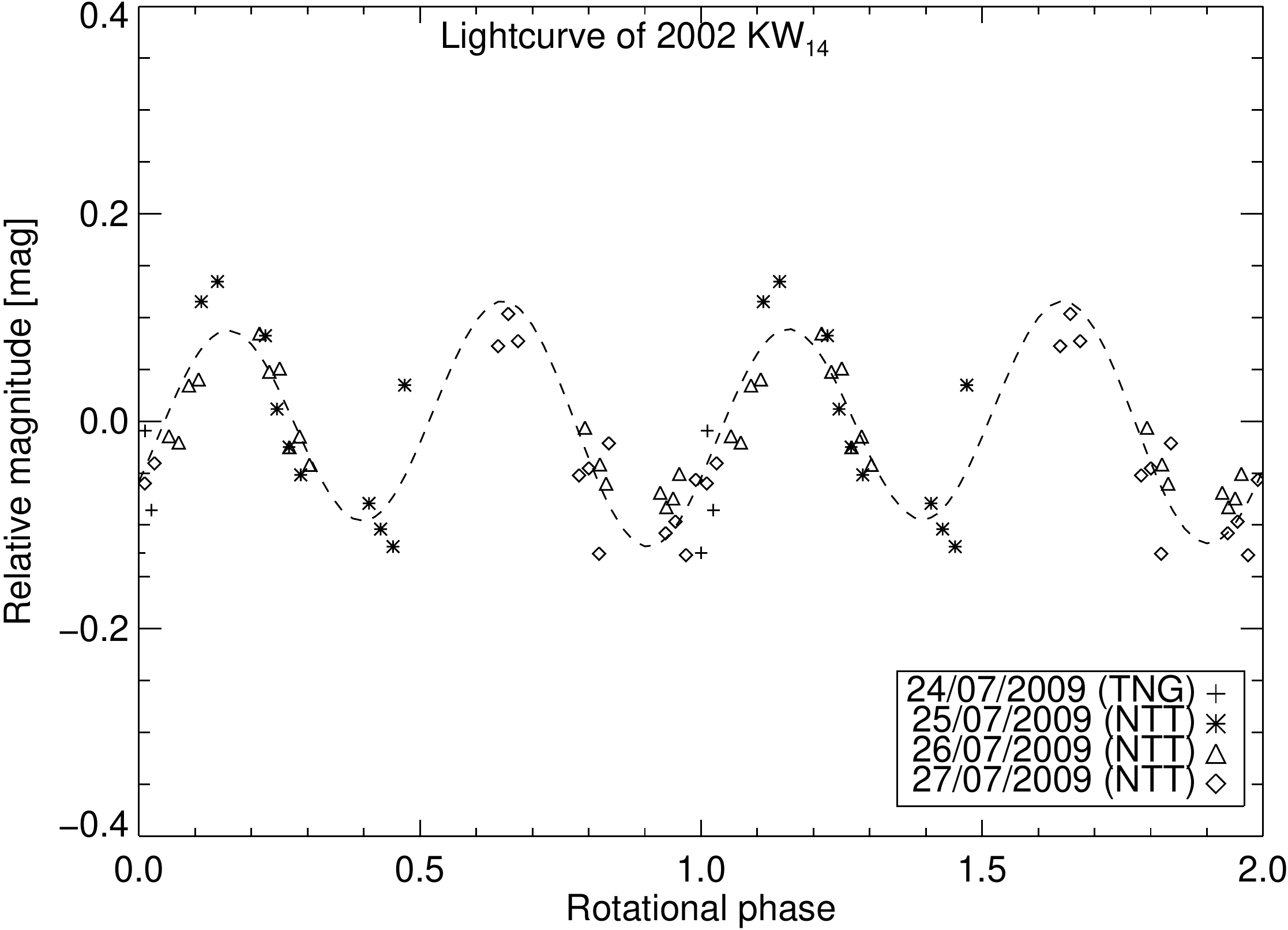}
\includegraphics[width=8cm, angle=90]{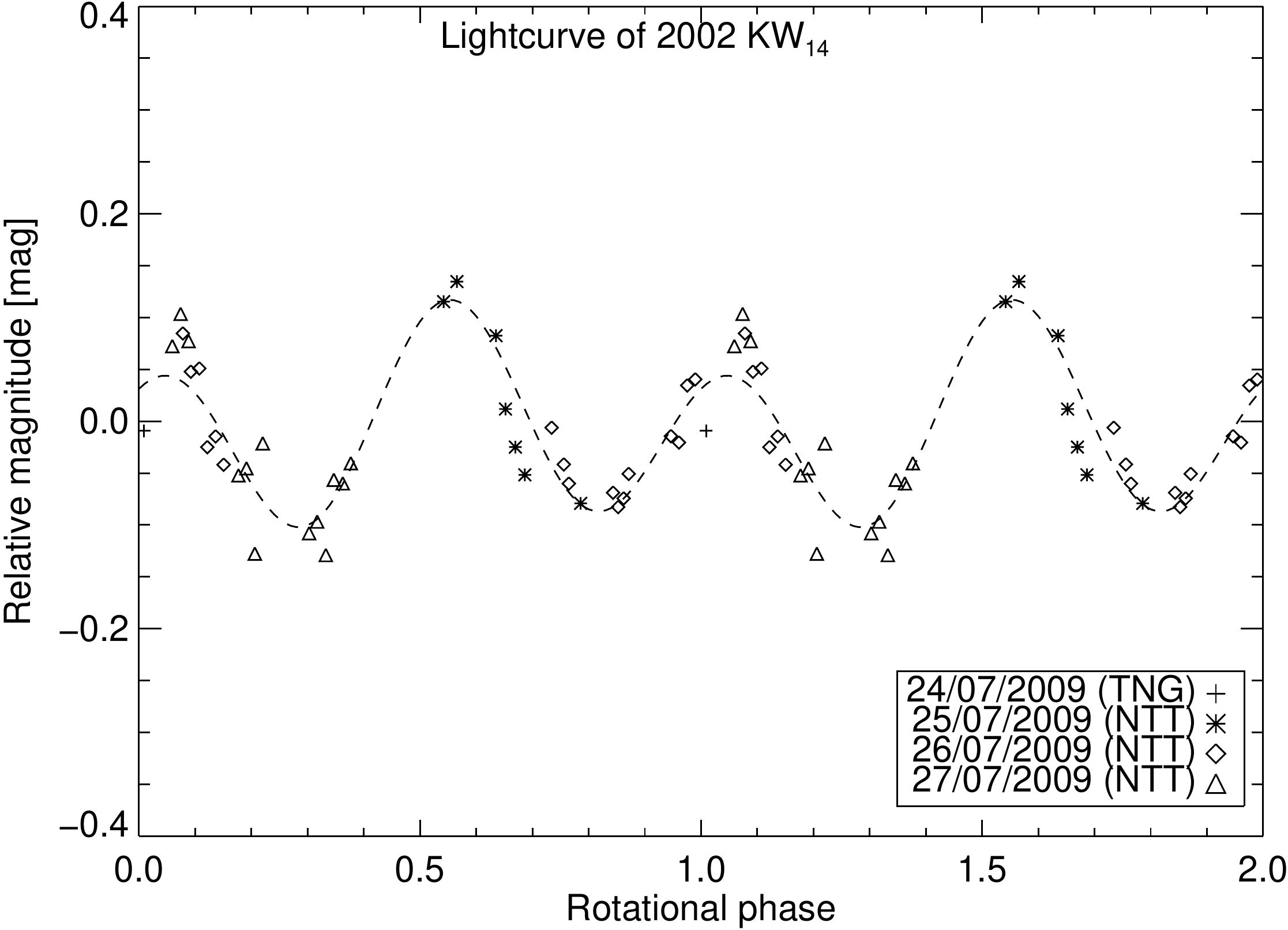}
\caption{Rotational phase curves for 2002~KW$_{14}$ obtained by using a spin
period of 8.58~h (upper plot) and 10.50~h (lower plot). The dash line is a Fourier Series fit of the photometric
data. Different symbols correspond to different dates.}
\label{fig6}
\end{center}
\end{figure*}

\clearpage

\begin{figure*}
\begin{center}
\includegraphics[width=8cm, angle=90]{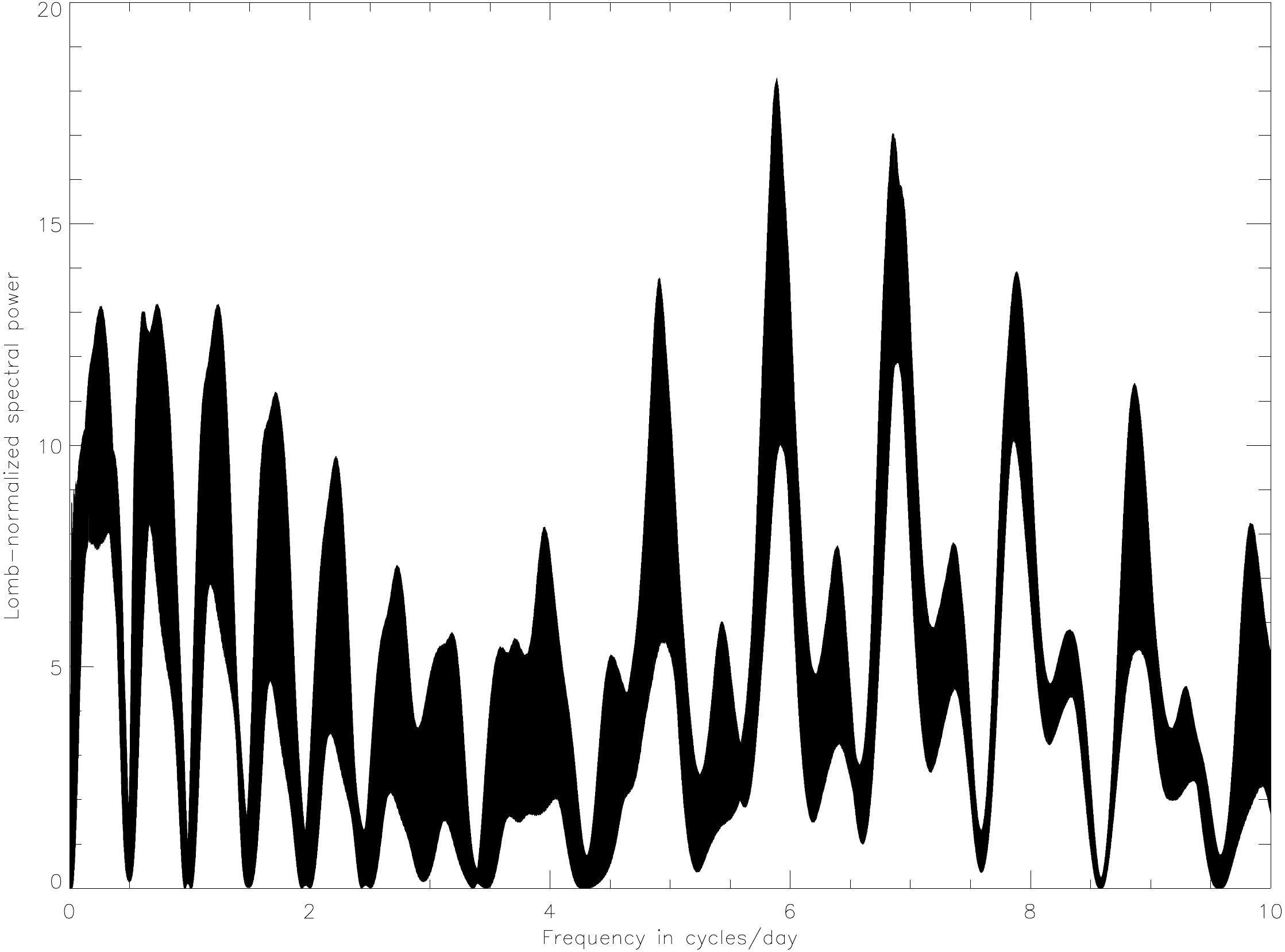}
\caption{Lomb periodogram of 2002~TX$_{300}$}
\label{fig7}
\end{center}
\end{figure*}
\begin{figure*}
\begin{center}
\includegraphics[width=8cm, angle=90]{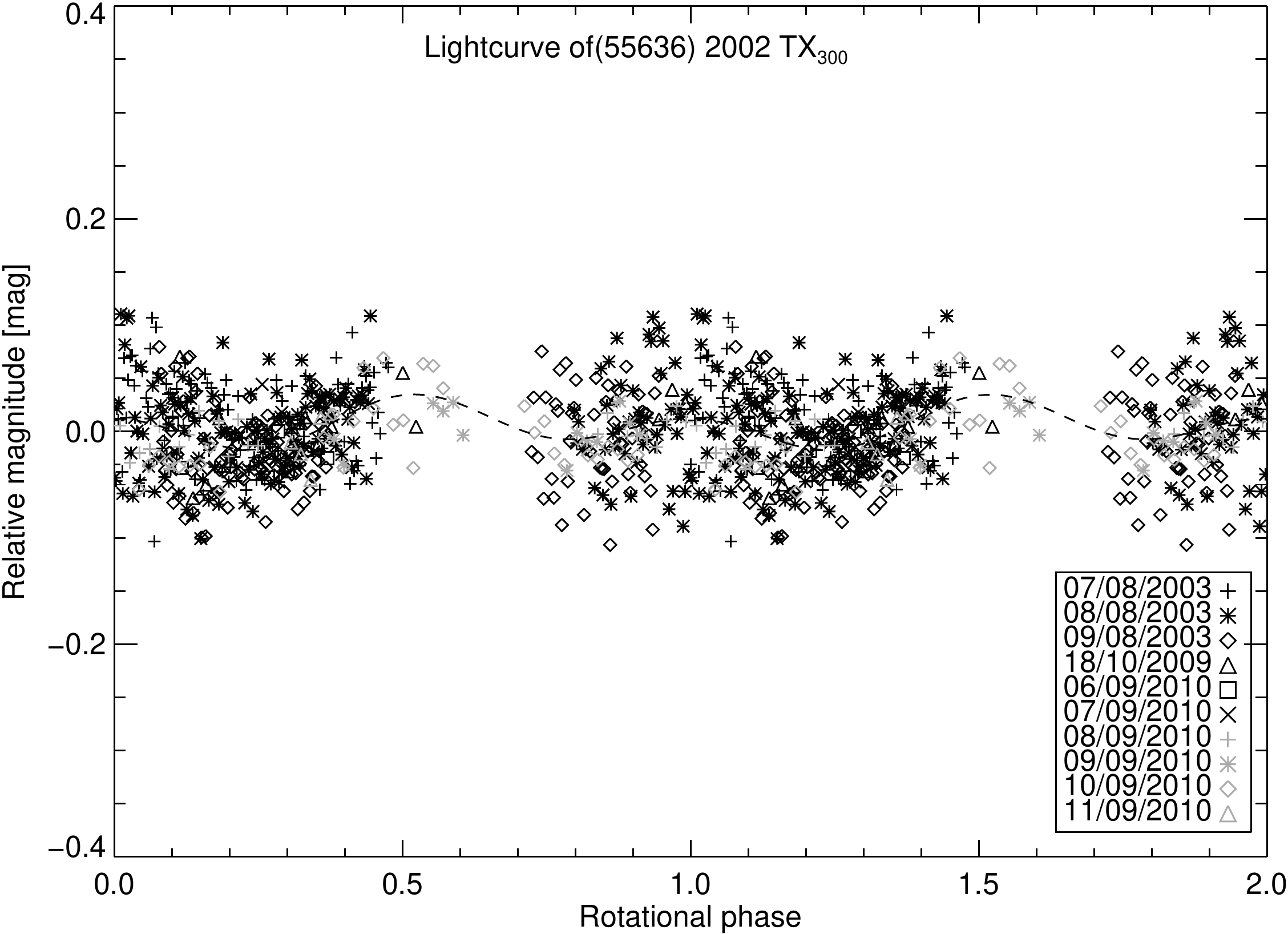}
\includegraphics[width=8cm, angle=90]{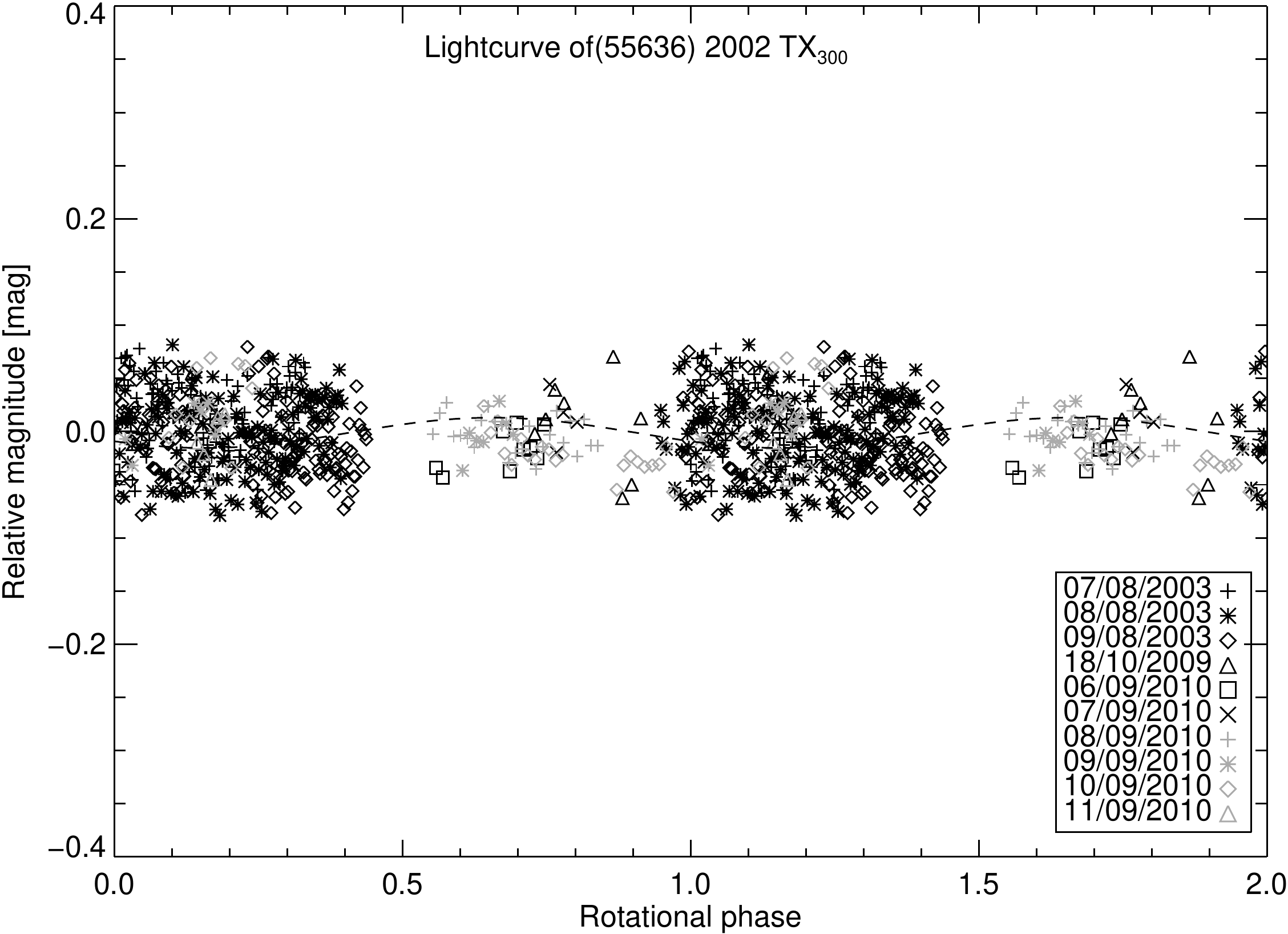}
\caption{Rotational phase curves for 2002~TX$_{300}$ obtained by using a spin
period of 8.15~h (upper plot) and 11.7~h (lower plot). The dash line is a Fourier Series fit of the photometric
data. Different symbols correspond to different dates.}
\label{fig8}
\end{center}
\end{figure*}

\clearpage

\begin{figure*}
\begin{center}
\includegraphics[width=8cm, angle=90]{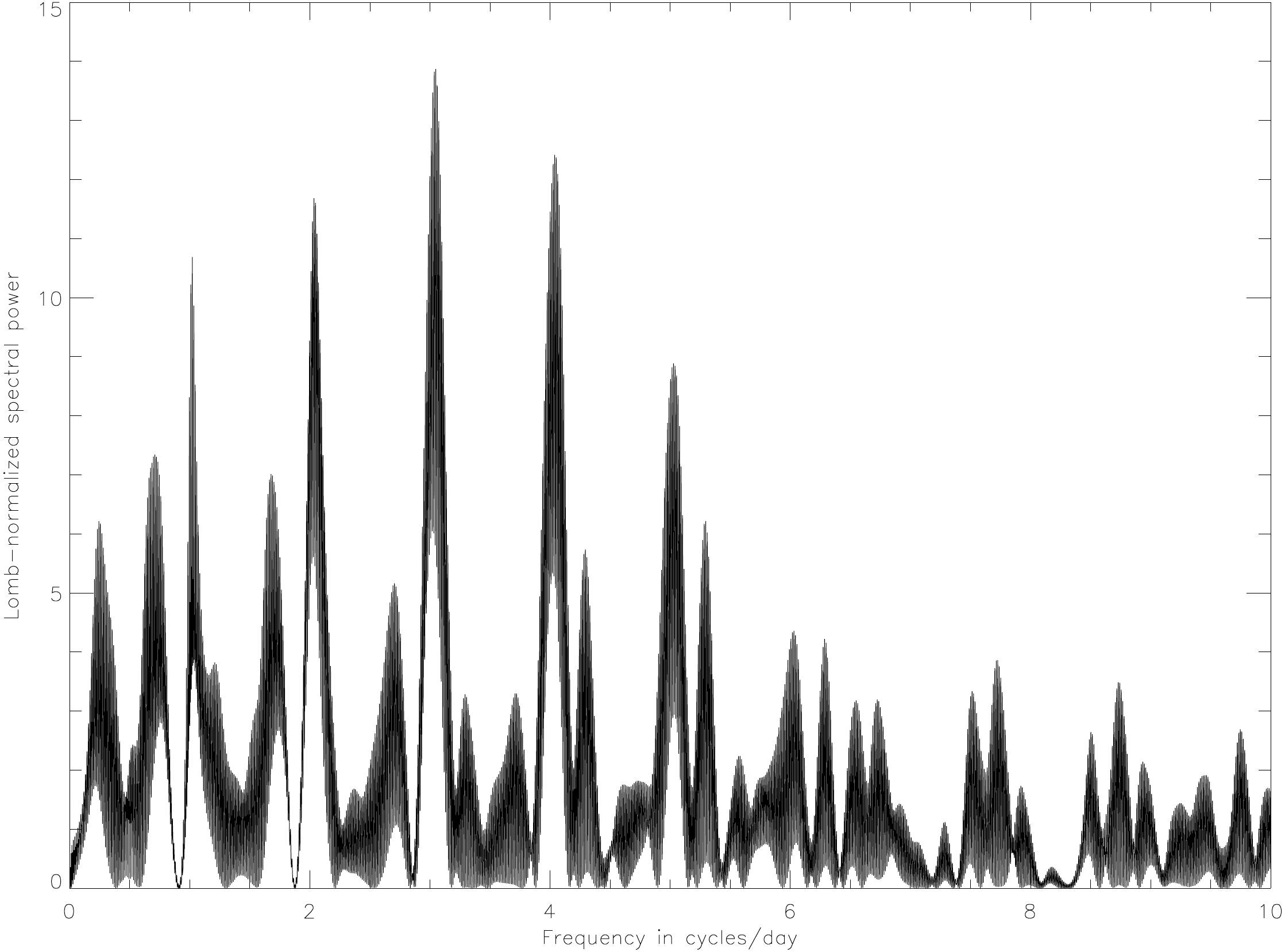}
\caption{Lomb periodogram of 2004~NT$_{33}$}
\label{fig9}
\end{center}
\end{figure*}
\begin{figure*}
\begin{center}
\includegraphics[width=8cm, angle=90]{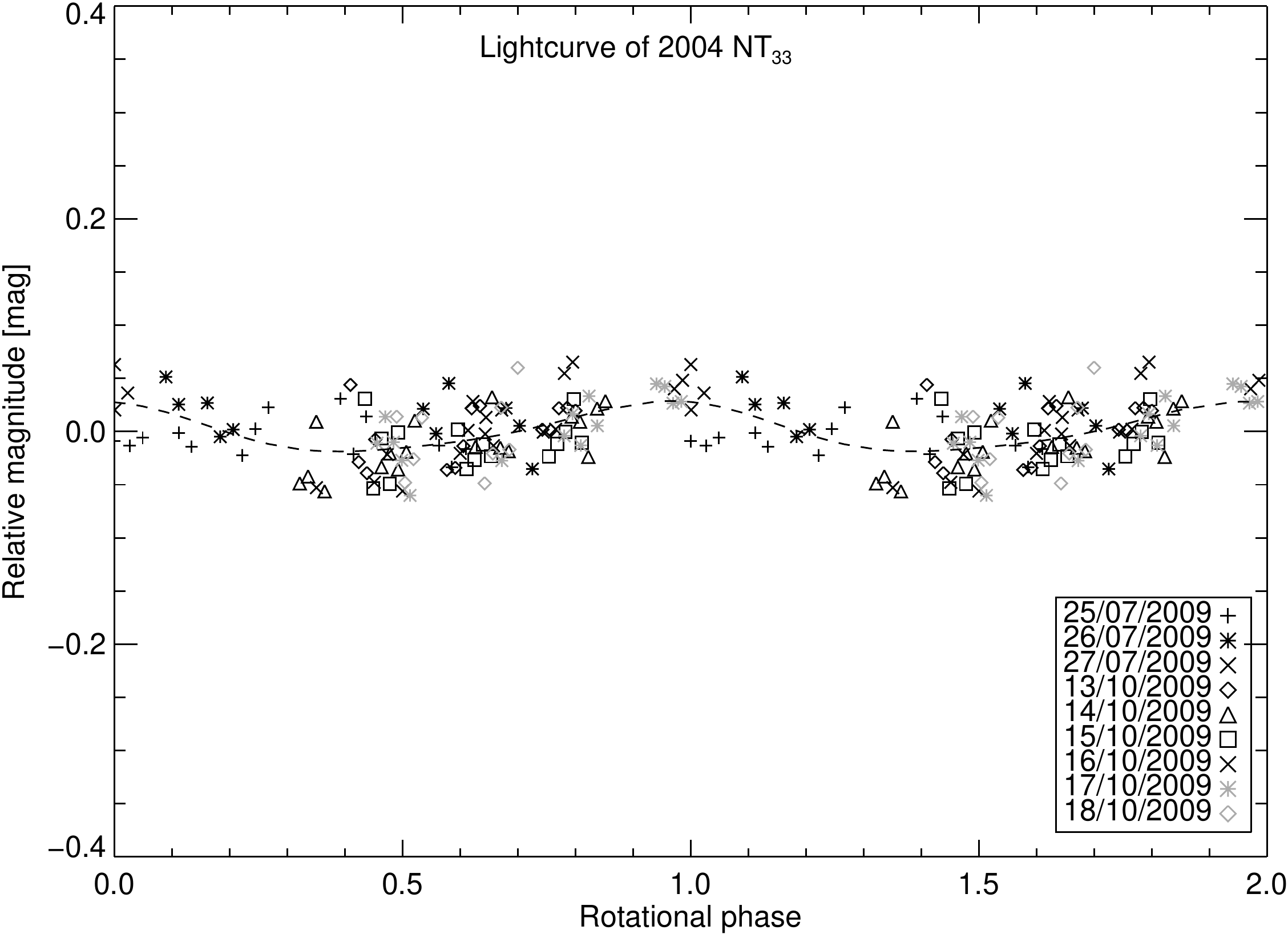}
\caption{Rotational phase curves for 2004~NT$_{33}$ obtained by using a spin
period of 7.87~h. The dash line is a Fourier Series fit of the photometric
data. Different symbols correspond to different dates.}
\label{fig10}
\end{center}
\end{figure*}

\clearpage

\begin{figure*}
\begin{center}
\includegraphics[width=8cm, angle=90]{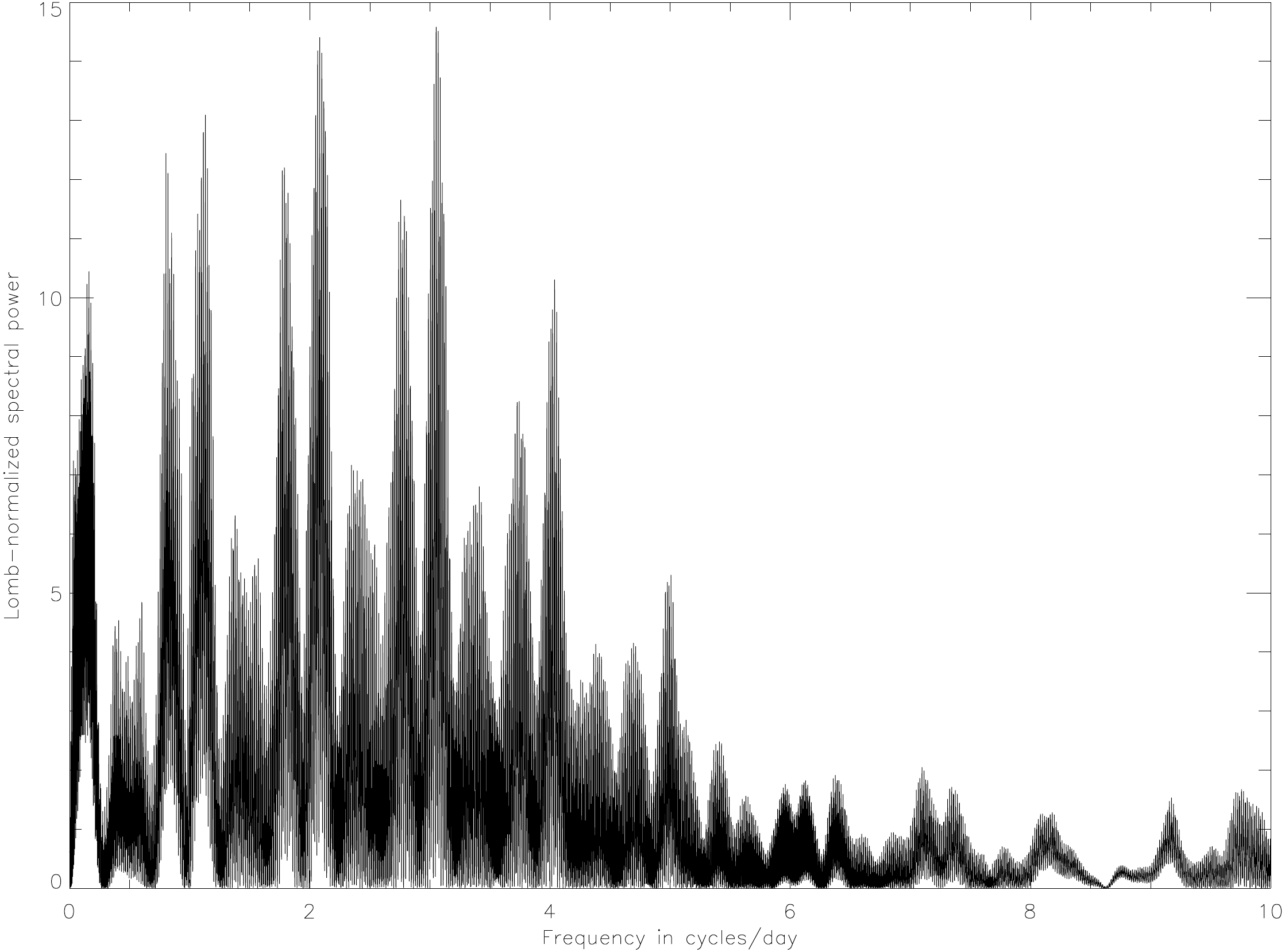}
\caption{Lomb periodogram of 2004~XA$_{192}$}
\label{fig11}
\end{center}
\end{figure*}
\begin{figure*}
\begin{center}
\includegraphics[width=8cm, angle=90]{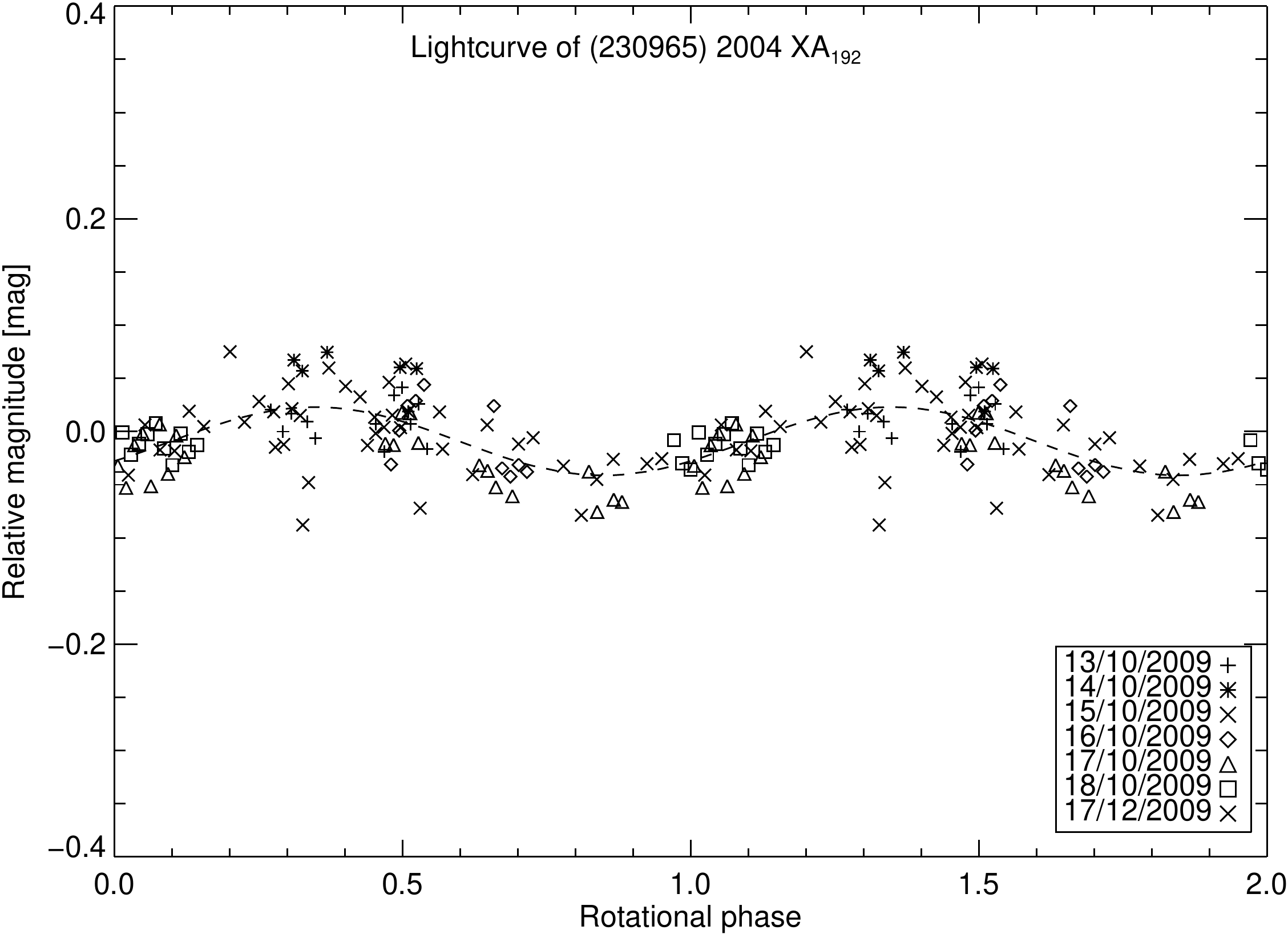}
\caption{Rotational phase curve for 2004~XA$_{192}$ obtained by using a
spin period of 7.88~h. The dash lines are a Fourier Series fits of
the photometric data. Different symbols correspond to different dates.}
\label{fig12}
\end{center}
\end{figure*}

\clearpage

\begin{figure*}
\begin{center}
\includegraphics[width=8cm, angle=90]{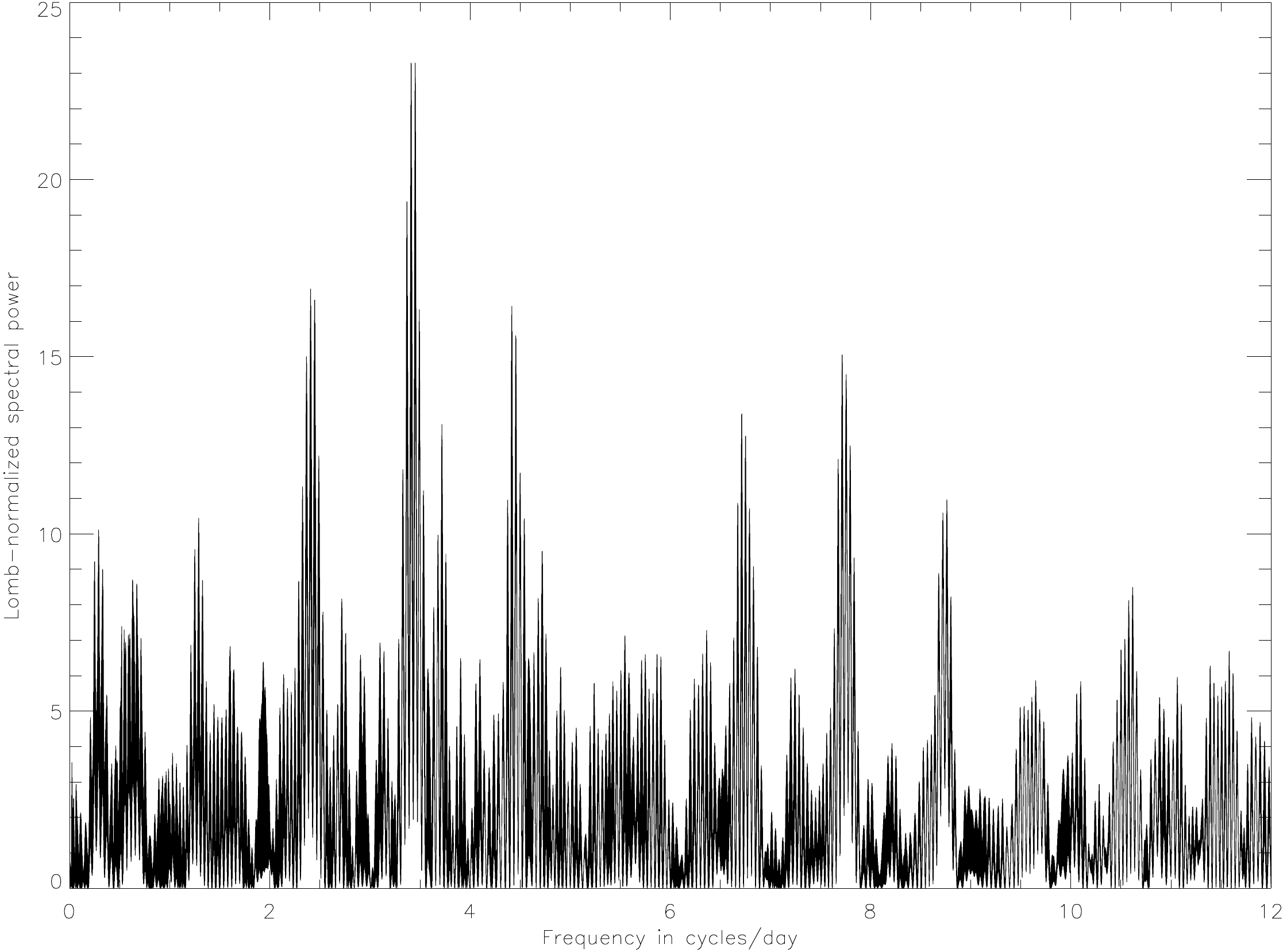}
\caption{Lomb periodogram of 2005~UQ$_{513}$}
\label{fig14}
\end{center}
\end{figure*}

\clearpage

\begin{figure*}
\begin{center}
\includegraphics[width=8cm, angle=90]{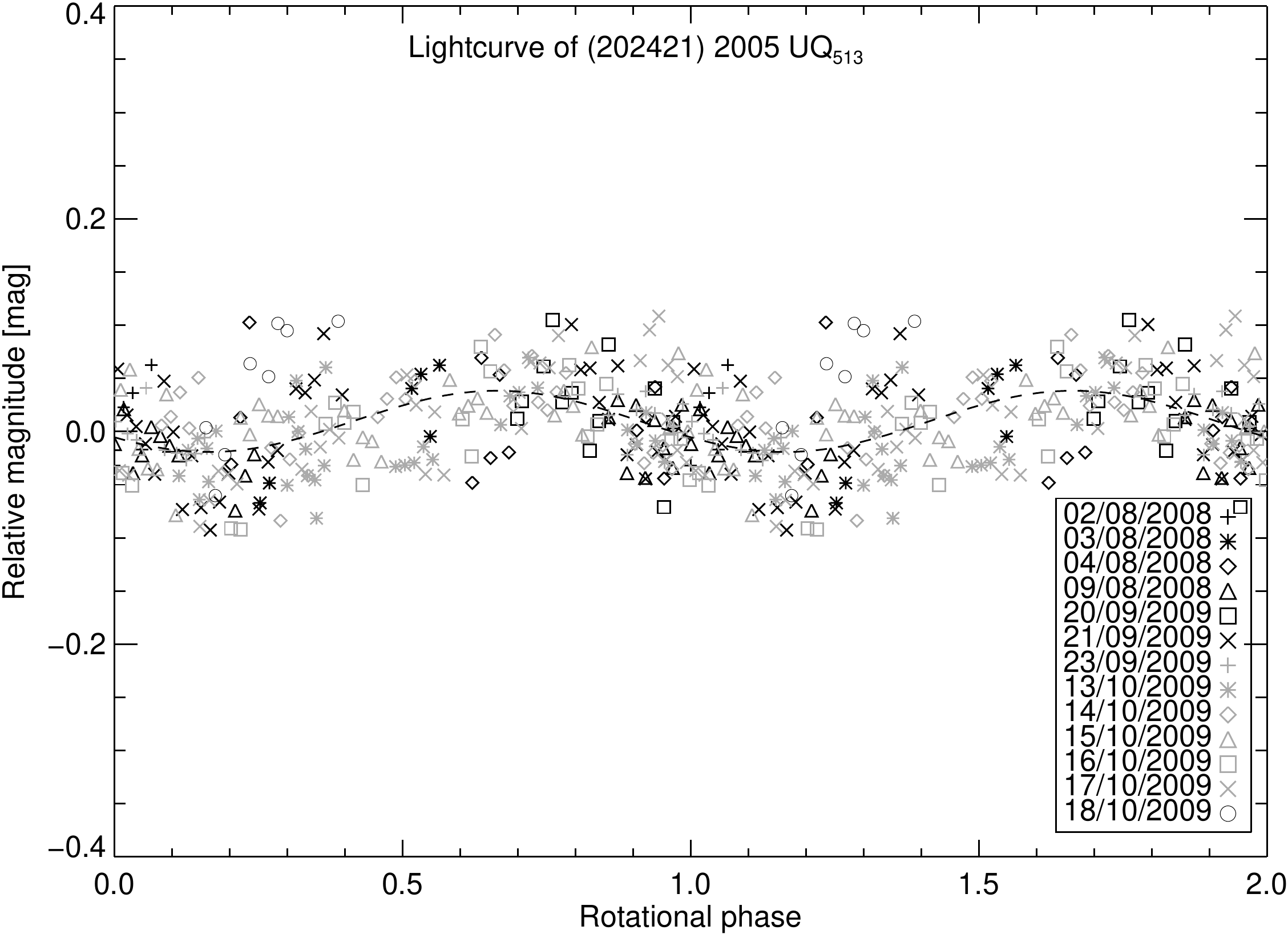}
\includegraphics[width=8cm, angle=90]{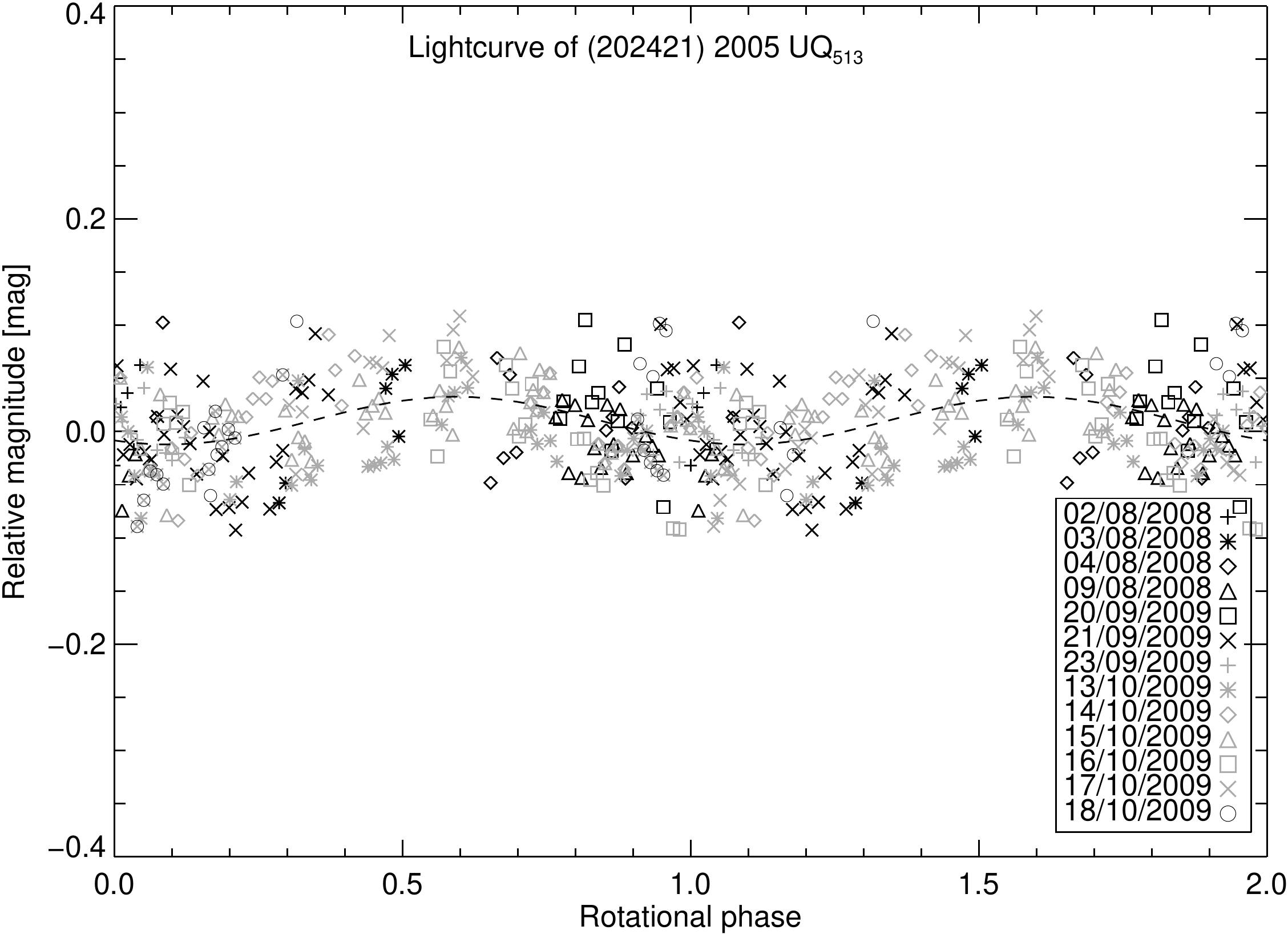}
\caption{Rotational phase curves for 2005~UQ$_{513}$ obtained by using a spin
period of 7.03~h (upper plot) and 10.01~h (lower plot). The dash line is a Fourier Series fit of the photometric
data. Different symbols correspond to different dates.}
\label{fig15}
\end{center}
\end{figure*}

\clearpage

\begin{figure*}
\begin{center}
\includegraphics[width=8cm, angle=90]{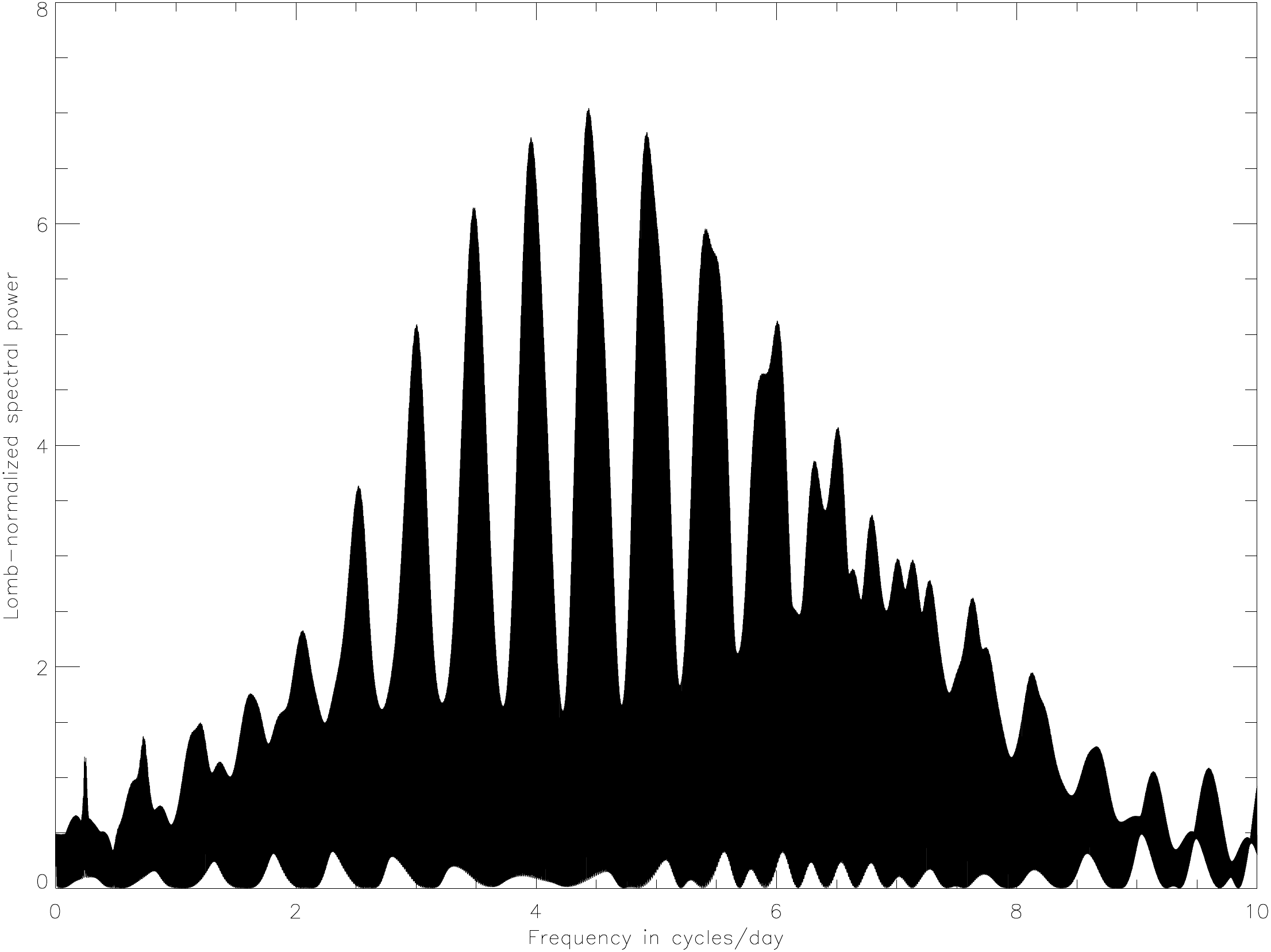}
\caption{Lomb periodogram of 2002~TC$_{302}$}
\label{fig16}
\end{center}
\end{figure*}
\begin{figure*}
\begin{center}
\includegraphics[width=8cm, angle=90]{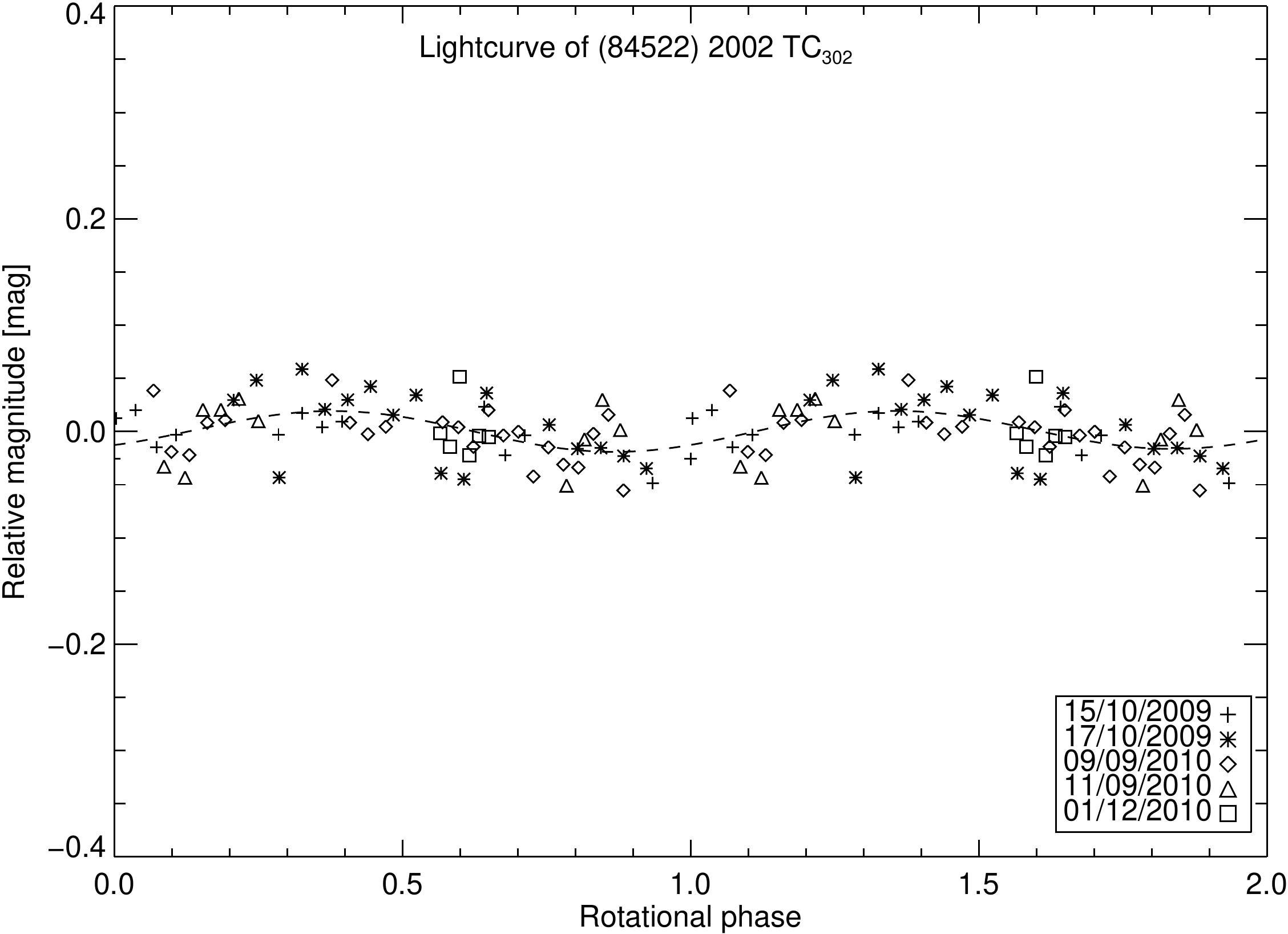}
\caption{Rotational phase curve for 2002~TC$_{302}$ obtained by using a spin
period of 5.41~h. The dash line is a Fourier Series fit of the photometric
data. Different symbols correspond to different dates.}
\label{fig17}
\end{center}
\end{figure*}

\clearpage

\begin{figure*}
\begin{center}
\includegraphics[width=8cm, angle=90]{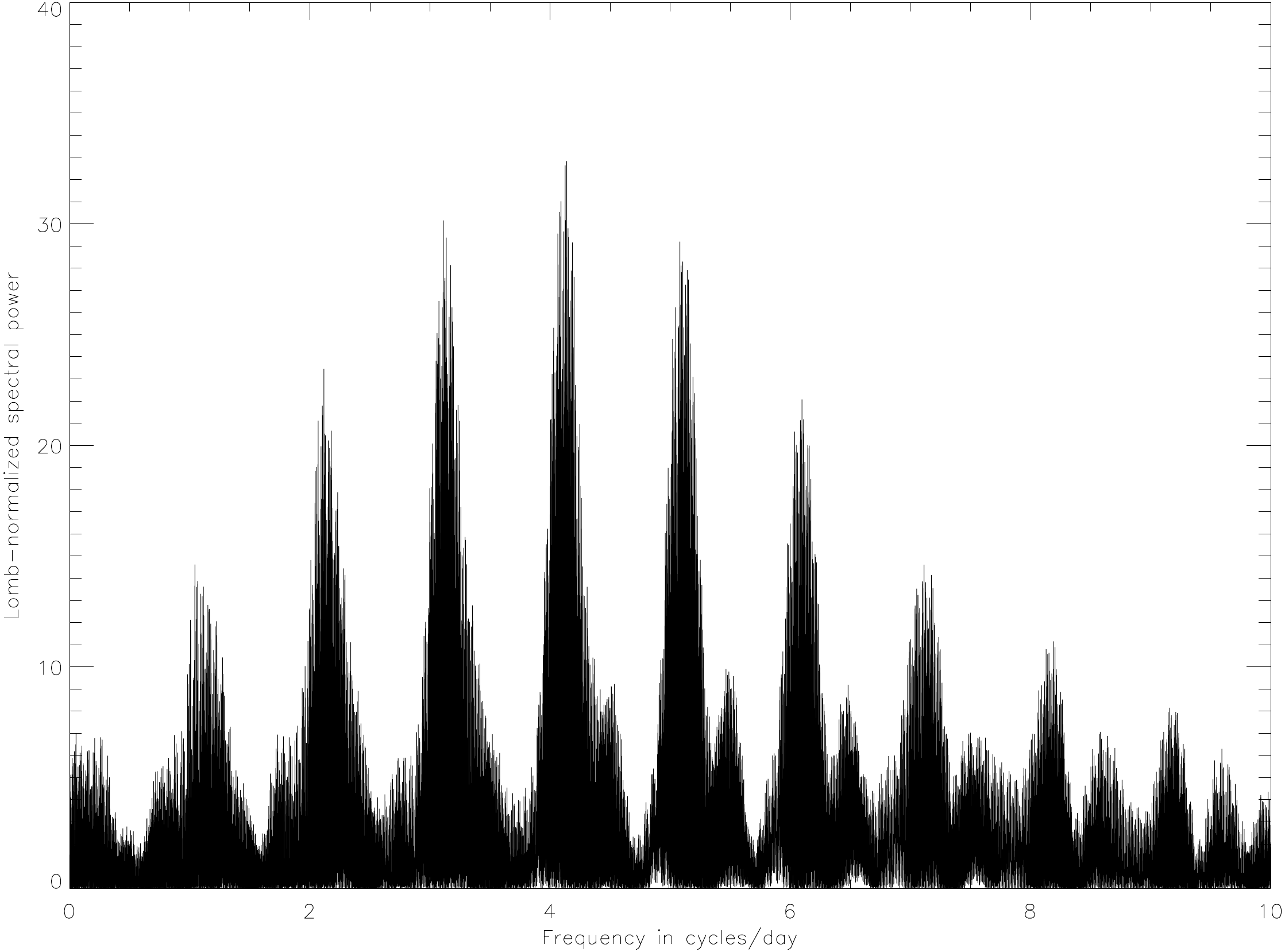}
\caption{Lomb periodogram of 1999~KR$_{16}$}
\label{fig18}
\end{center}
\end{figure*}

\begin{figure*}
\begin{center}
\includegraphics[width=8cm, angle=90]{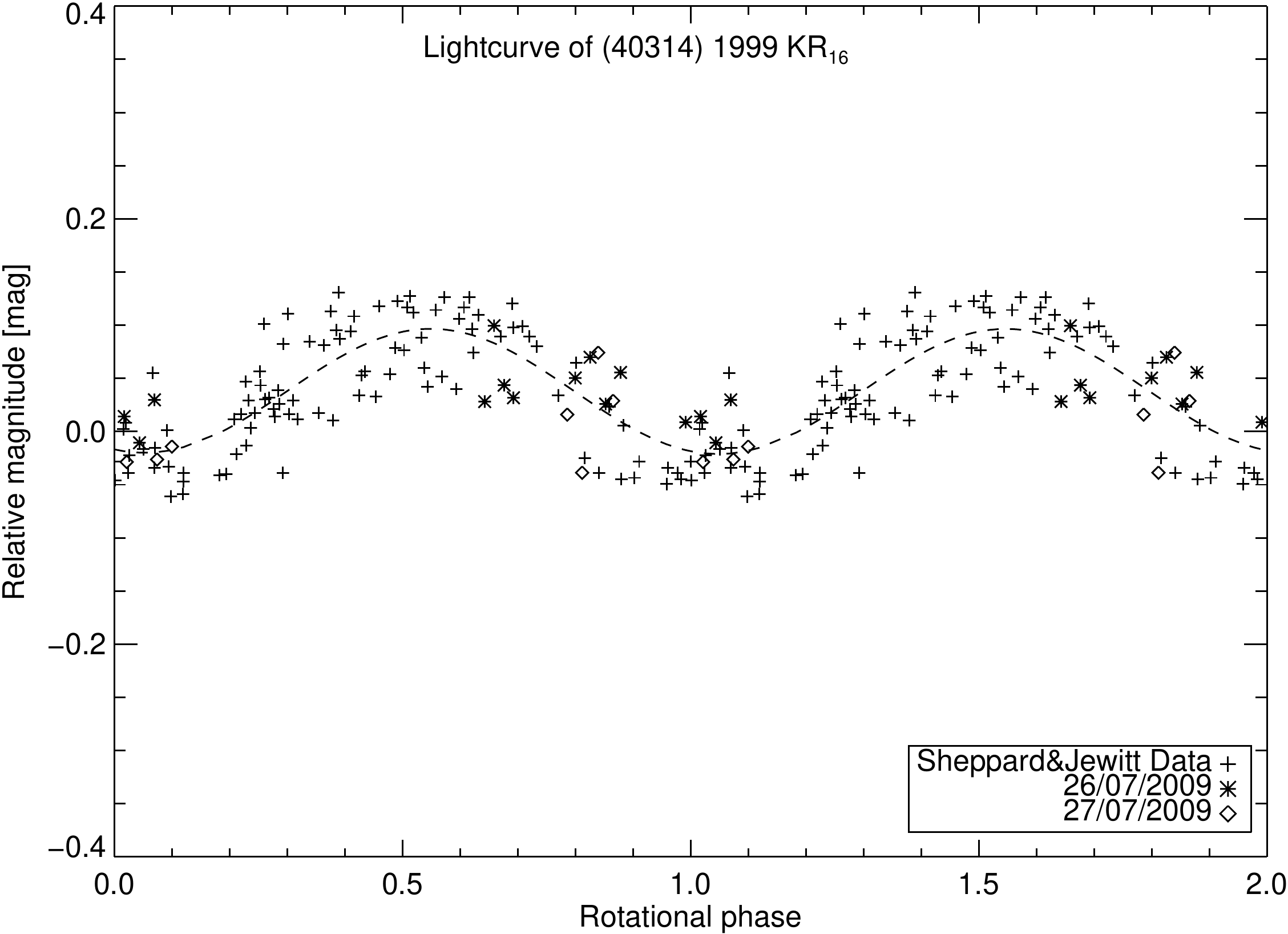}
\caption{Rotational phase curve for 1999~KR$_{16}$ obtained by using a spin
period of 5.8~h. The dash line is a Fourier Series fit of the photometric
data. Different symbols correspond to different dates.}
\label{fig19}
\end{center}
\end{figure*}

\clearpage

\begin{figure*}
\begin{center}
\includegraphics[width=8cm, angle=90]{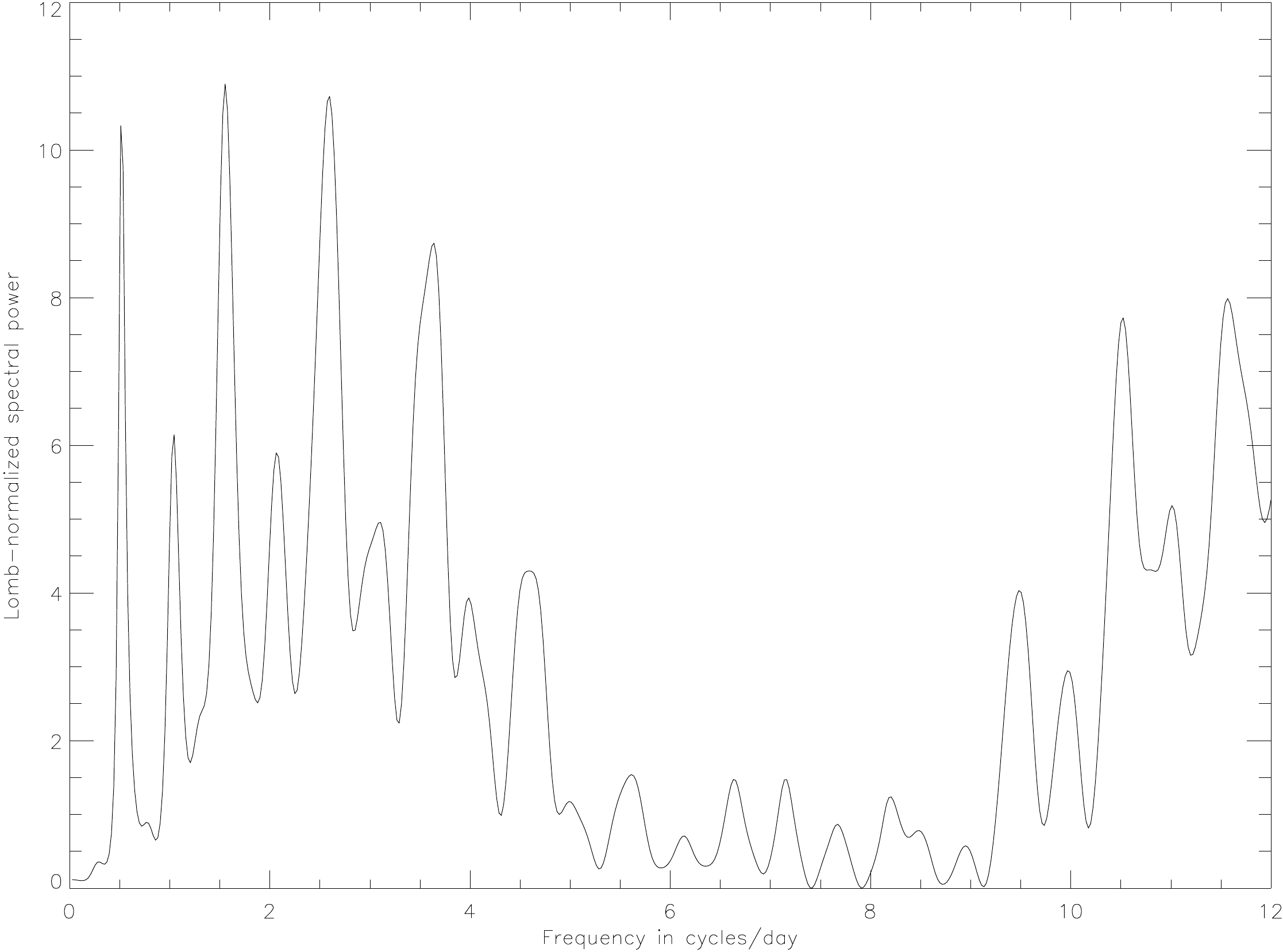}
\caption{Lomb periodogram of 1999~OX$_{3}$}
\label{fig20}
\end{center}
\end{figure*}

\clearpage

\begin{figure*}
\begin{center}
\includegraphics[width=8cm, angle=90]{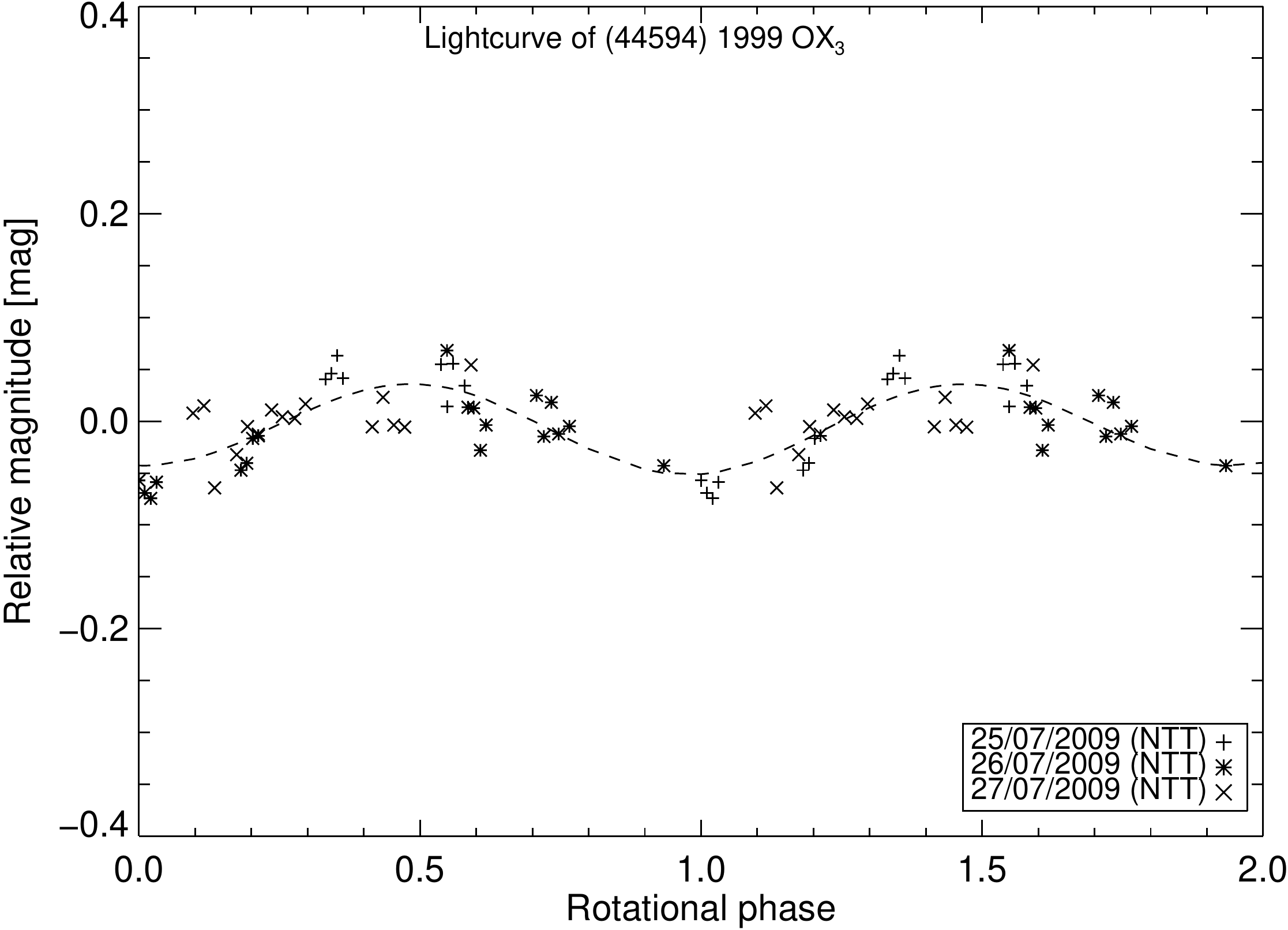}
\includegraphics[width=8cm, angle=90]{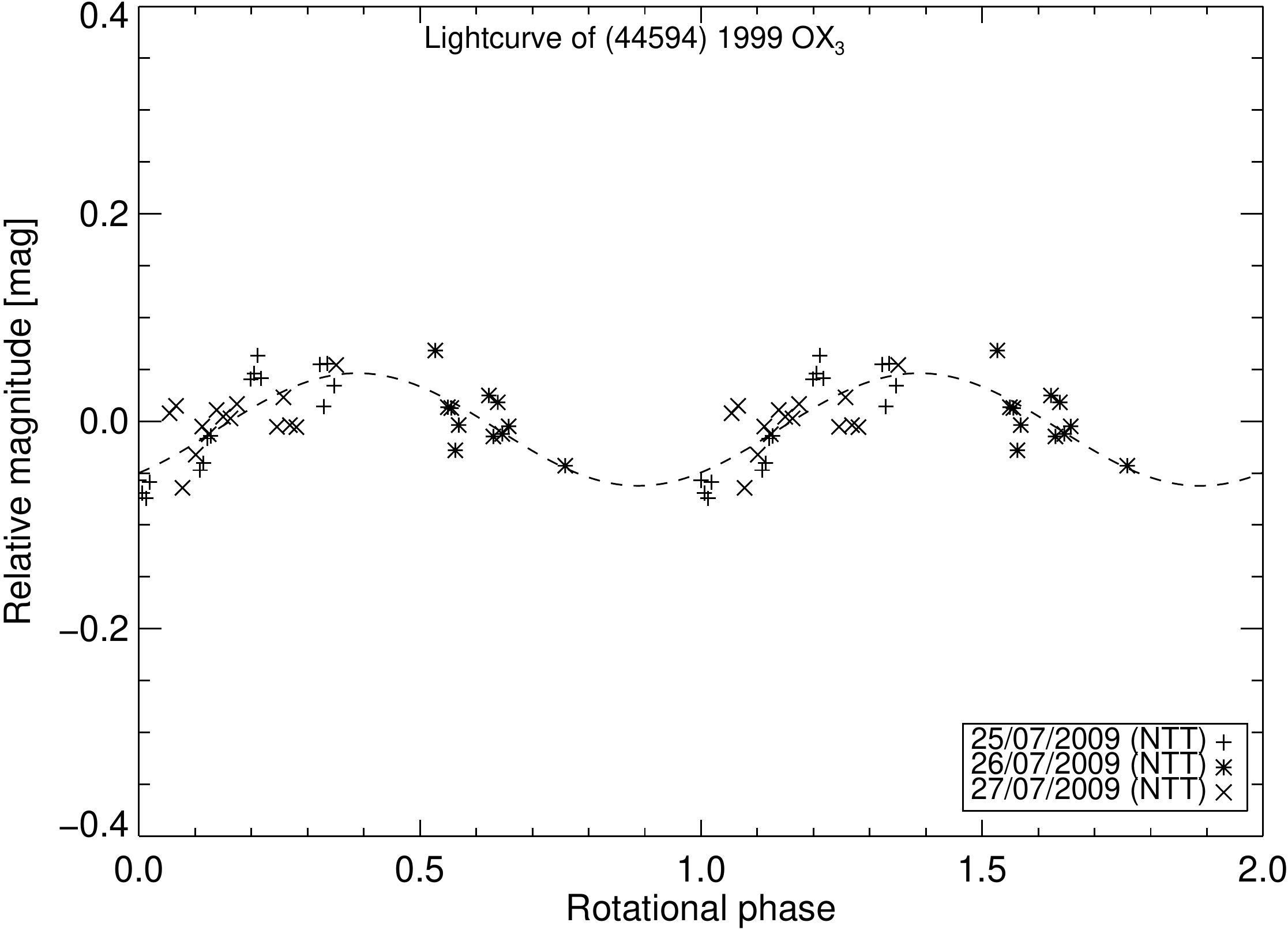}
\caption{Rotational phase curves for 1999~OX$_{3}$ obtained by using different
spin periods; 9.26~h (upper plot) and 15.45~h (lower plot). In both cases, we
present a single peak lightcurve. The dash lines are a Fourier Series fits of
the photometric data. Different symbols correspond to different dates.}
\label{fig21}
\end{center}
\end{figure*}

\clearpage

\begin{figure*}
\begin{center}
\includegraphics[width=8cm, angle=90]{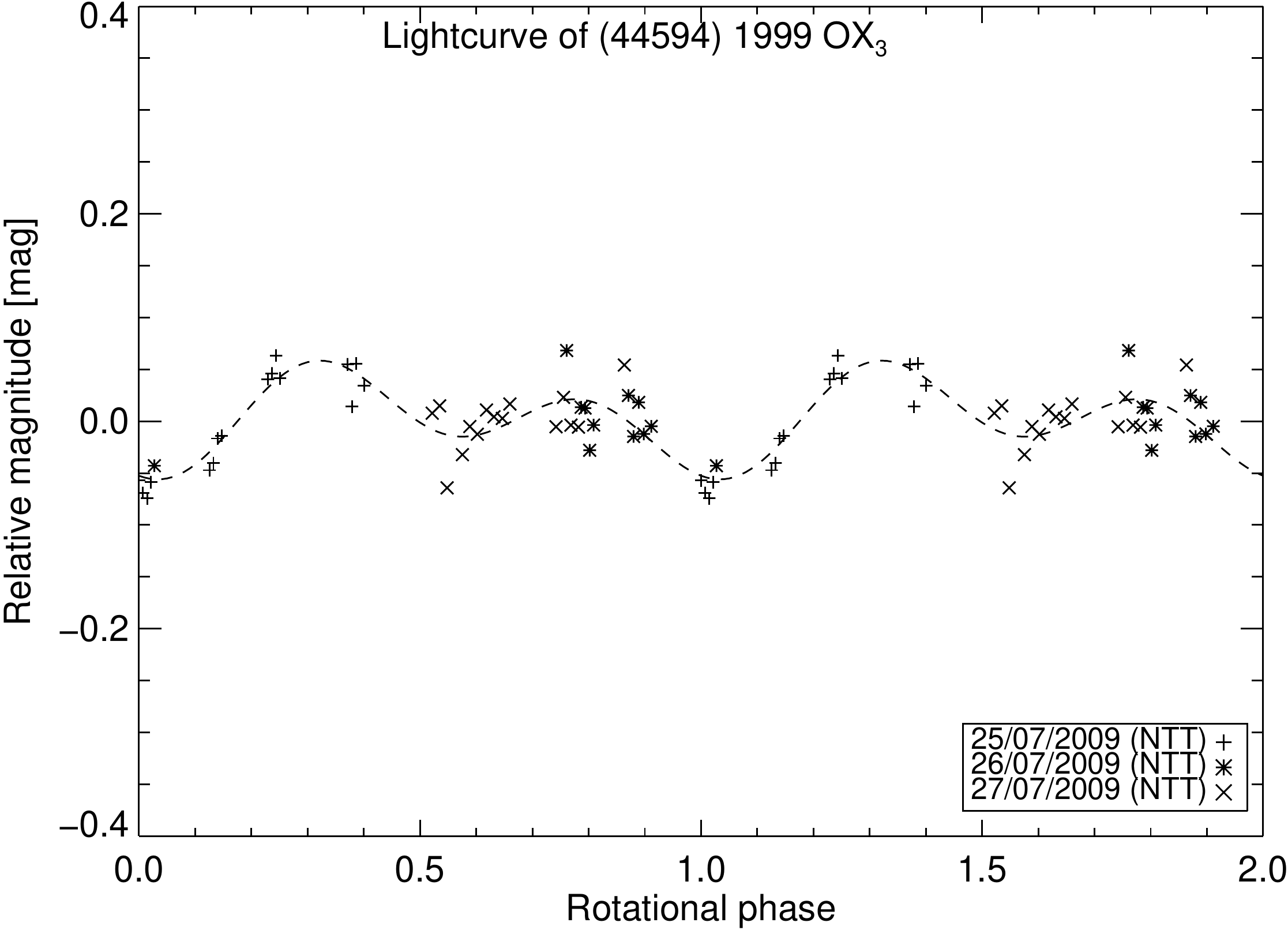}
\caption{Rotational phase curves for 1999~OX$_{3}$ obtained by using a rotational period of 13.4~h. The dash lines are a Fourier Series fits of
the photometric data. Different symbols correspond to different dates.}
\label{fig21}
\end{center}
\end{figure*}

\clearpage

\begin{figure*}
\begin{center}
\includegraphics[width=8cm, angle=90]{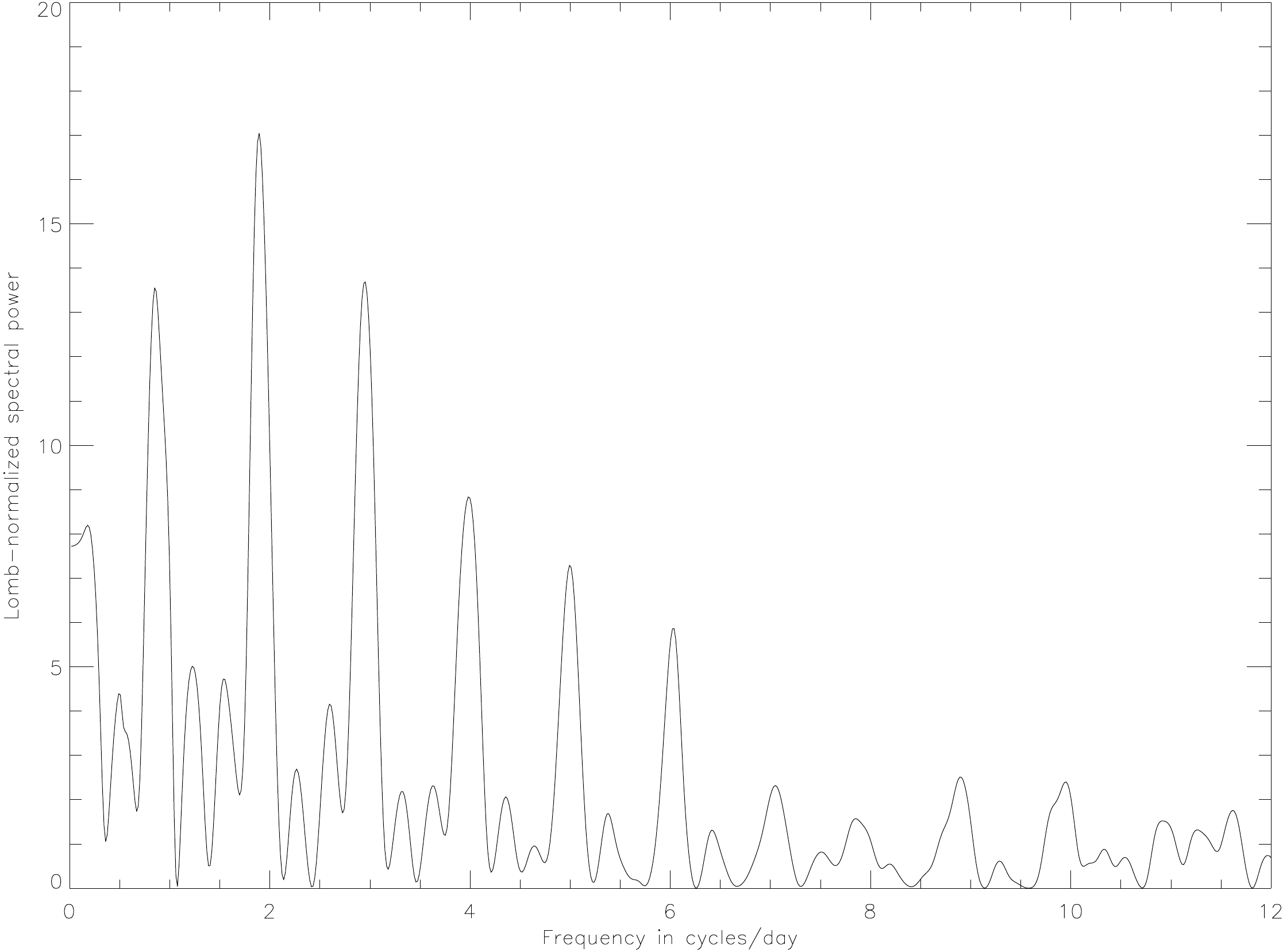}
\caption{Lomb periodogram of 2005~TB$_{190}$}
\label{fig22}
\end{center}
\end{figure*}

\begin{figure*}
\begin{center}
\includegraphics[width=8cm, angle=90]{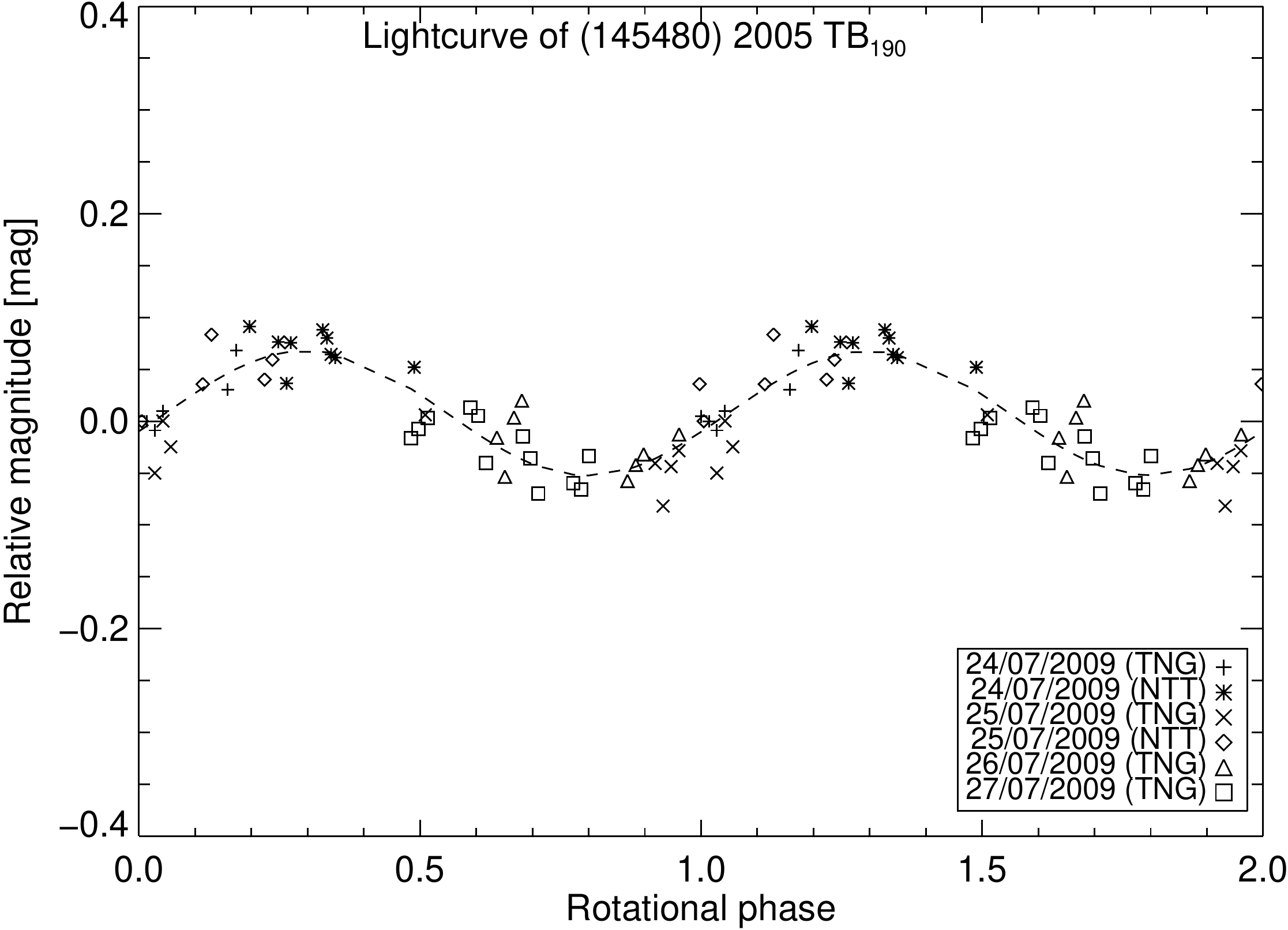}
\caption{Rotational phase curve for 2005~TB$_{190}$ obtained by using a
spin period of 12.68~h. The dash lines are a Fourier Series fits of
the photometric data. Different symbols correspond to different dates.}
\label{fig23}
\end{center}
\end{figure*}

\clearpage

\begin{figure*}
\begin{center}
\includegraphics[width=8cm, angle=90]{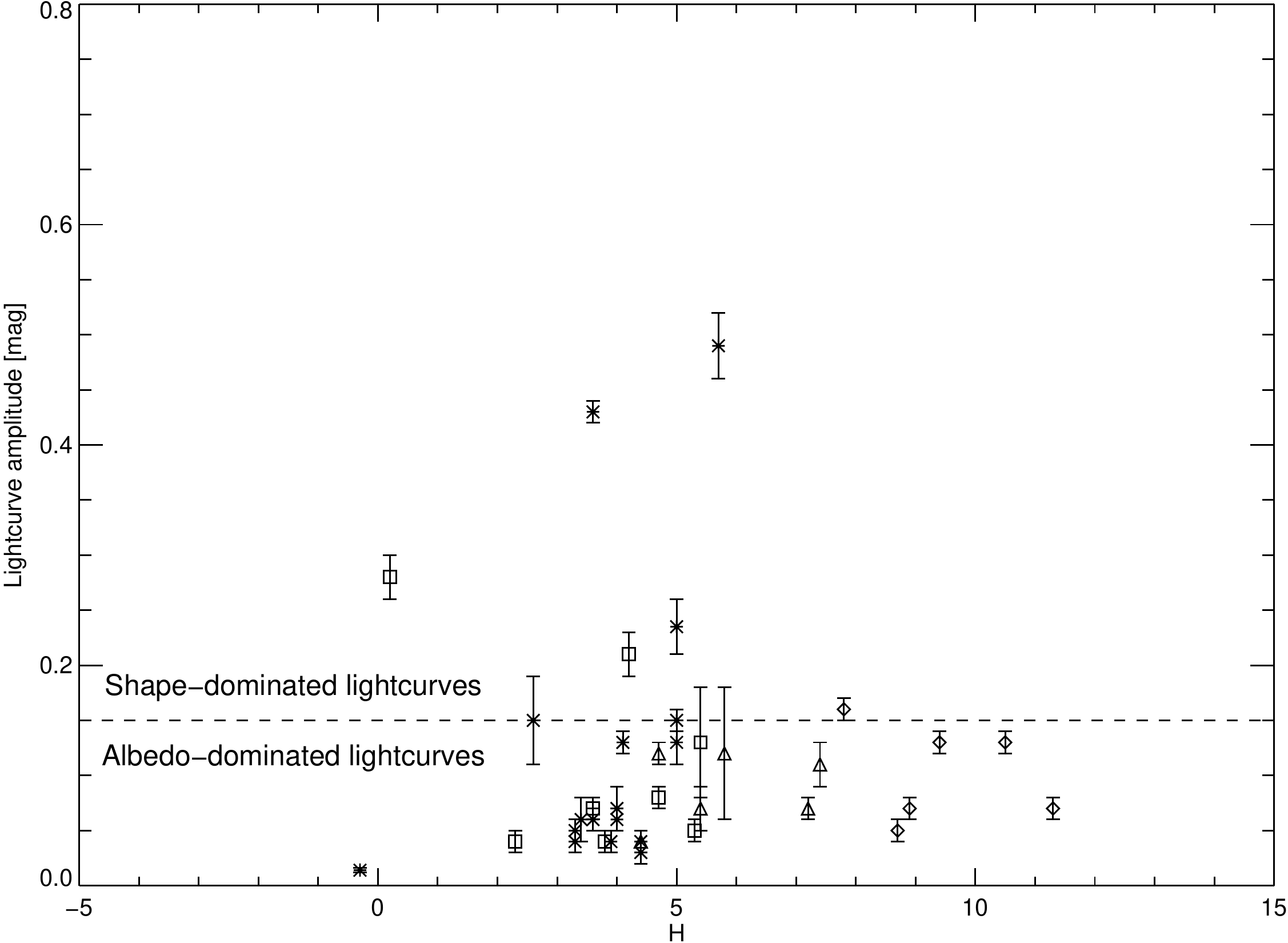}
\caption{Lightcurve Amplitude vs. Absolute Magnitude: All objects presented in this work and in Thirouin et al. (2010) are plotted: squares for Resonant Objects, asterisks for Classical Objects, triangles for Scattered and Detached disk Objects and diamonds for Centaurs. As mentioned in the discussion section, the sample of studied objects is highly biased toward bright objects and we note the lack of lightcurve with high amplitude. In fact, except, cases like 2001~QY$_{297}$, Varuna or Haumea, majority of studied objects present a low amplitude. Line at 0.15~mag is indicating the separation between the shape and albedo dominated lightcurves. Absolute magnitudes extracted from the MPC database.}
\label{fig22}
\end{center}
\end{figure*}
\begin{figure*}
\begin{center}
\includegraphics[width=8cm, angle=90]{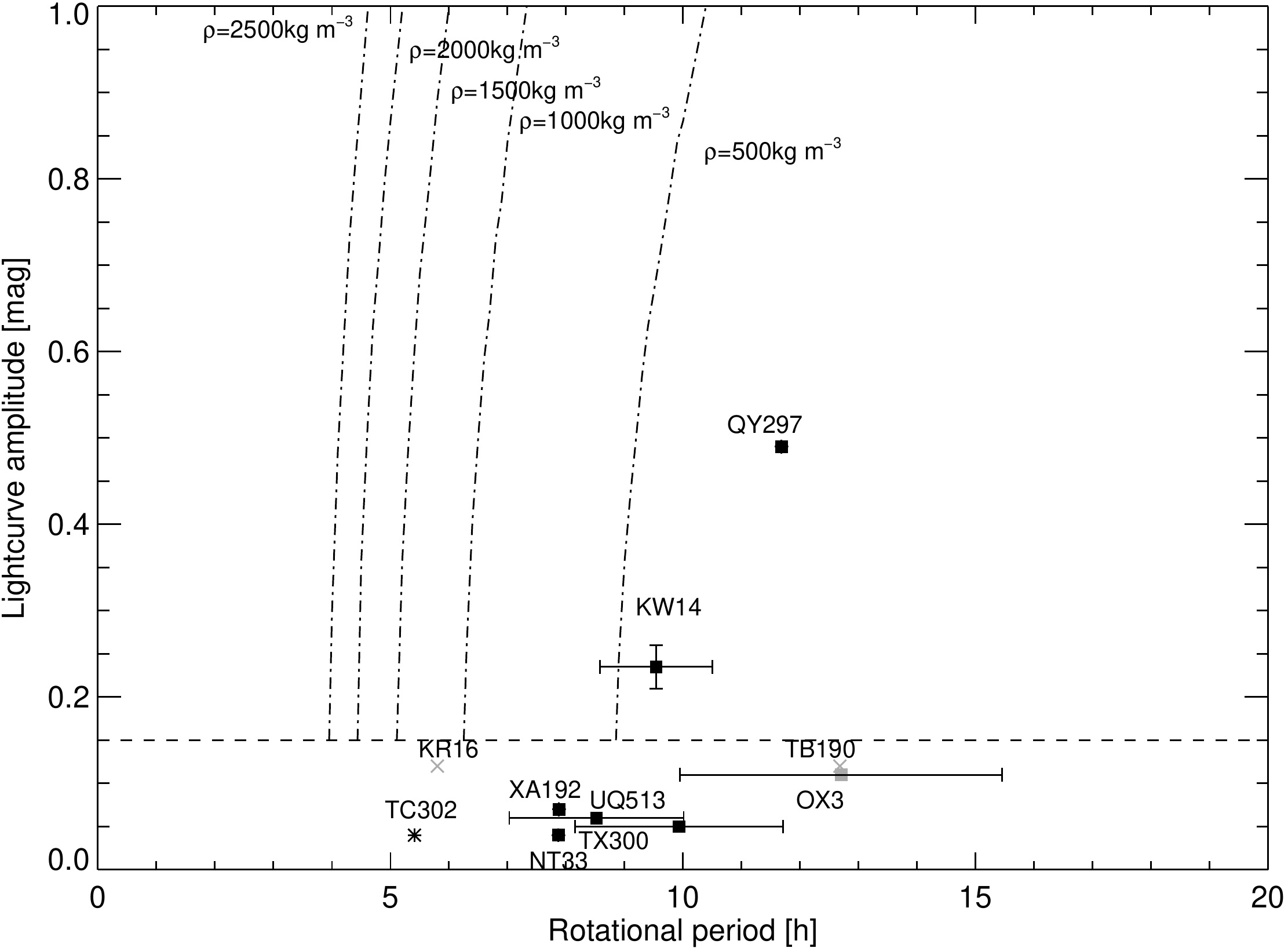}
\caption{Lightcurve Amplitude vs. Rotational Period for theoretical Jacobi ellipsoids of various densities compared with observations. All objects presented in this work are shown: black crosses for Resonant Objects, black squares for Classical Objects, gray squares for Scattered disk Objects and gray cross for Detached Objects. For each target, we indicate the last part of its name. For example, 2001~QY$_{297}$ is indicated as QY297. In the case of various rotational periods are found for the same target, we plot the average value and the corresponding error bars. Horizontal line defines the separation between shape and albedo dominated lightcurves as in the previous plot. 
Each vertical dash line defines a density value. Density values are indicated on the top of each line. This plot is updated from Duffard et al. (2009) in which a complete explanation of the plot can be found. This study assumes that TNOs are in hydrostatic equilibrium.}
\label{fig24}
\end{center}
\end{figure*}

\clearpage

\begin{figure*}
\begin{center}
\includegraphics[width=8cm, angle=90]{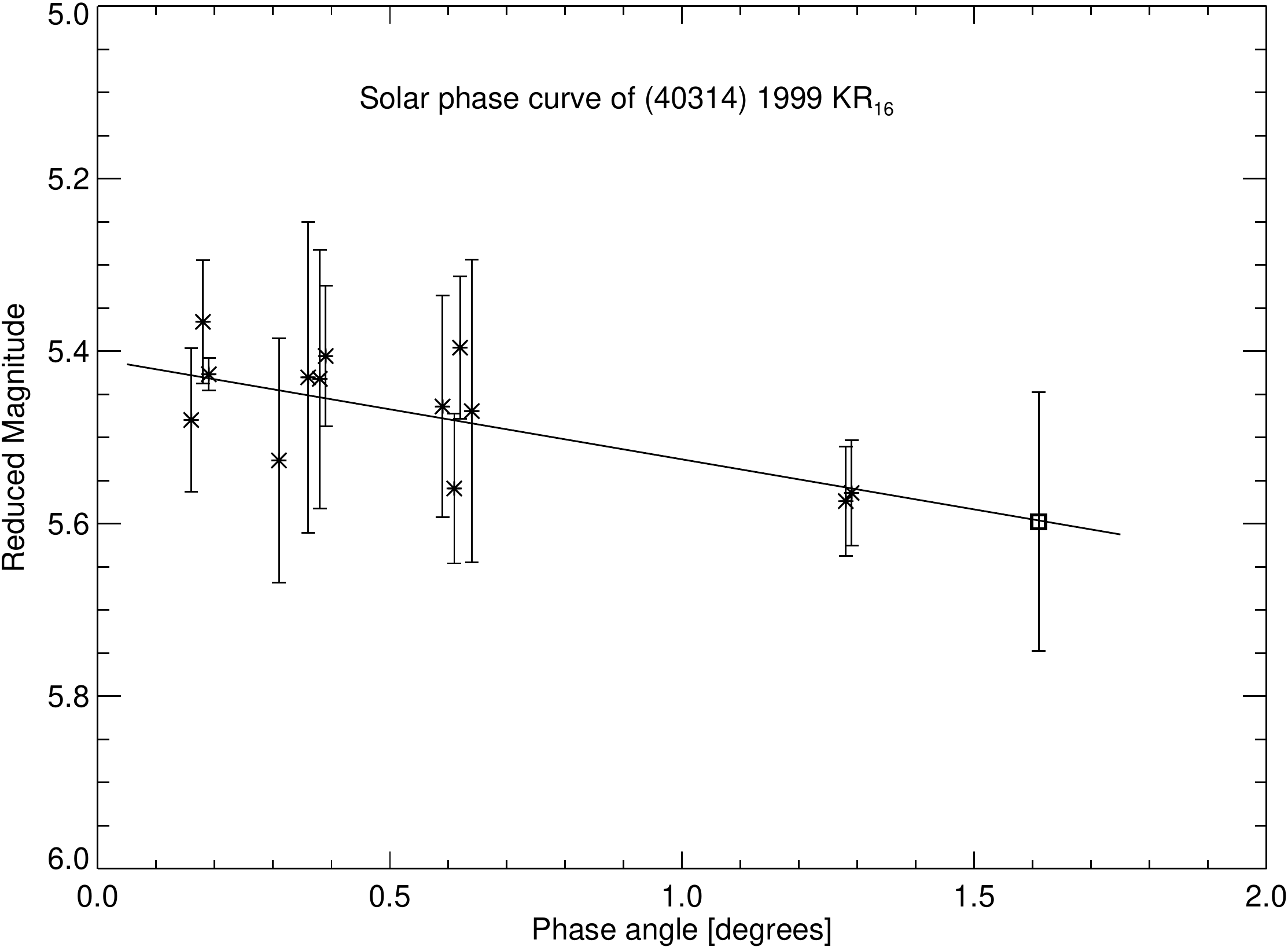}
\caption{Reduced Magnitude vs. Phase Angle for (40314) 1999~KR$_{16}$: we plot data published in Sheppard $\&$ Jewiit (2002) with an asterisk symbol and data reported in this work with a square symbol. Continuous line is a linear fit of all data.  
}
\label{fig25}
\end{center}
\end{figure*}
\begin{figure*}
\begin{center}
\includegraphics[width=8cm, angle=90]{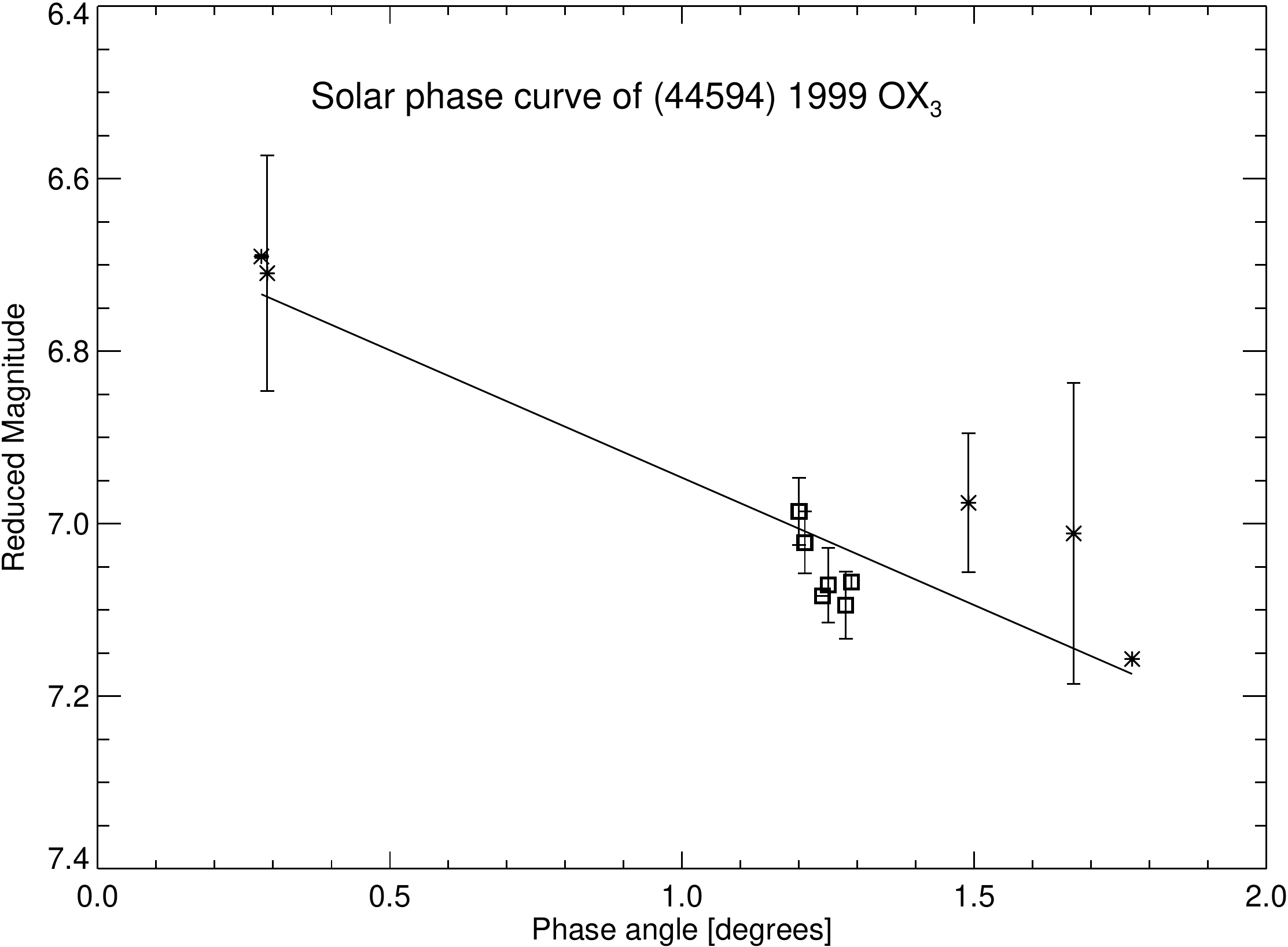}
\caption{Reduced Magnitude vs. Phase Angle for (44594) 1999~OX$_{3}$: we plot data published in Bauer et al. (2003) with an asterisk symbol and data reported in this work with a square symbol. Continuous line is a linear fit of all data.  
}
\label{fig26}
\end{center}
\end{figure*}

\clearpage

\begin{scriptsize}
\begin{onecolumn}
\begin{longtable}{cccccccccc}
\caption{This full table is available online. We present our photometric results: the name of the object and for each image we specify the Julian date (not corrected for light time), the Relative magnitude [mag] and the 1-$\sigma$ error associated [mag], the R magnitude [mag], the filter used during observational runs, the phase angle ($\alpha$) [deg], topocentric (r$_{h}$) and heliocentric ($\Delta$) distances [AU] and the magnitude [mag] at 1~AU from the Earth and at 1~AU from the Sun. We highlight in bold face the date of the image in which we performed a crude absolute calibration (see Data Reduction section).}\\

Object  & Julian date  & Relative magnitude & Error & R magnitude & Filter  & $\alpha$ & r$_\mathrm{h}$ & $\Delta$ & m$_\mathrm{R}$(1,1) \\ 
\hline
(40314) 1999~KR$_{16}$ &	2455039.46074	&	0.028	&	0.019	&	21.19	&	R	&	1.61	&	35.913	&	36.034	&	5.63	\\
&	2455039.46466	&	0.100	&	0.018	&	21.24	&	R	&	1.61	&	35.913	&	36.034	&	5.68	\\
&	2455039.46887	&	0.044	&	0.018	&	21.31	&	R	&	1.61	&	35.913	&	36.034	&	5.75	\\
&	2455039.47279	&	0.032	&	0.016	&	21.31	&	R	&	1.61	&	35.913	&	36.034	&	5.75	\\
&	2455039.49869	&	0.050	&	0.027	&	21.20	&	R	&	1.61	&	35.913	&	36.034	&	5.64	\\
&	2455039.50493	&	0.070	&	0.031	&	21.21	&	R	&	1.61	&	35.913	&	36.034	&	5.65	\\
 
\end{longtable}
\end{onecolumn}
\end{scriptsize}

\end{document}